\documentclass{LMCS}

\usepackage{array}
\usepackage{epic, eepic}
\usepackage{amssymb,amsmath,stmaryrd}
\usepackage[ruled, vlined]{algorithm2e}
\usepackage{slashbox}

\usepackage{gastex,hyperref,enumerate}

\newcounter{compressEnum}
\renewcommand{\thecompressEnum}{$\roman{compressEnum}$}
\newenvironment{compressEnum}
{\setcounter{compressEnum}{0}}% \begin{paragraph}}

{}%\end{paragraph}}

\newcommand{\itCompress}{\stepcounter{compressEnum}{(\thecompressEnum) }}

\makeatletter

\let\c@lemm\c@theo
\let\c@coro\c@theo
\let\c@defi\c@theo
\let\c@assu\c@theo

\makeatother

\def\sg{\mathrel[\joinrel\mathrel[}
\def\sd{\mathrel]\joinrel\mathrel]}

\def\abs#1{\ensuremath{\lvert #1\rvert}}
\def\norm#1{\ensuremath{\lVert #1\rVert}}

\newcommand{\Pre}{\mathsf{Pre}}

\newcommand{\nat}{\mathbb N}

\newcommand{\real}{{\mathbb R}}

\newcommand{\closure}[1]{\downarrow \!\!#1}
\newcommand{\sem}[1]{\sg \mathrel{#1} \sd}

\renewcommand{\l}{{\ell}}

\newcommand{\Loc}{{\sf Loc}}

\newcommand{\tuple}[1]{\langle #1 \rangle}

\renewcommand{\inf}{\mathsf{inf}}

\newcommand{\A}{\mathcal{A}}
\newcommand{\B}{\mathcal{B}}
\newcommand{\F}{\mathcal{F}}

\renewcommand{\L}{\mathcal{L}}

\newcommand{\DAG}{{\sc dag}}

\newcommand{\odd}{\mathsf{odd}}

\newcommand{\ceilOdd}[1]{\lceil \mathrel {#1} \rceil^{\mathsf{odd}}}
\newcommand{\ceilEven}[1]{\lceil \mathrel {#1} \rceil^{\mathsf{ even}}}
\newcommand{\true}{{\sf true}}
\newcommand{\false}{{\sf false}}

\newcommand{\MH}{{\sf MH}}
\newcommand{\KV}{{\sf KV}}
\newcommand{\KVMH}{{\sf KVMH}}
\newcommand{\Max}{{\sf Max}}

\def\doi{5 (1:5) 2009}
\lmcsheading%
{\doi}
{1--20}
{}
{}
{Aug.~13, 2007}
{Mar.~\phantom{0}2, 2009}
{}   

\begin{document}

\title[]{Antichains for the Automata-Based Approach to Model-Checking\rsuper*}

\author[L.~Doyen]{Laurent Doyen\rsuper a} \address{{\lsuper a}CCS,
  \'{E}cole Polytechnique F\'ed\'erale de Lausanne, Switzerland}
\email{laurent.doyen@epfl.ch}

%\thanks{}

\author[J.-F.~Raskin]{Jean-Fran\c{c}ois Raskin\rsuper b}	%optional
\address{{\lsuper b}CS, Universit\'{e} Libre de Bruxelles, Belgium}	%optional
\email{jraskin@ulb.ac.be}  %optional

\titlecomment{{\lsuper *}%
  A preliminary version of this paper appeared in the
  \emph{Proceedings of the 13th International Conference on Tools and
    Algorithms for the Construction and Analysis of Systems} (TACAS),
  Lecture Notes in Computer Science 4424, Springer-Verlag, 2007,
  pp. 451-465.\\\indent This research was supported in part by the FRFC
  project ``Centre F\'ed\'er\'e en V\'erification'' funded by the
  Belgian National Science Foundation (FNRS) under grant nr 2.4530.02,
  by the PAI program Moves supported by the Belgian Federal
  Gouvernment: ``Fundamental Issues in Modelling, Verification and
  Evolution of Software'' ({\tt http://moves.vub.ac.be}), by the Swiss
  National Science Foundation, and by the European COMBEST project.  }

% Section 1

\begin{abstract}
  We propose and evaluate antichain algorithms to solve the universality and
  language inclusion problems for nondeterministic B\"uchi automata,
  and the emptiness problem for alternating B\"uchi automata.
  To obtain those algorithms, we establish the existence of
  simulation pre-orders that can be exploited to efficiently evaluate fixed
  points on the automata defined during the complementation step (that
  we keep implicit in our approach). We evaluate the performance of the
  algorithm to check the universality of B\"uchi automata
  using the random automaton model recently proposed by
  Tabakov and Vardi.  We show that on the difficult instances of this
  probabilistic model, our algorithm outperforms the standard ones by
  several orders of magnitude. 
  % This work is an extension to the
  % infinite words case of new algorithms for the finite words case
  % that we and co-authors have presented in a recent paper~\cite{cav06}.
\end{abstract} 

\maketitle

\section{Introduction}

In the automata-based approach to model-checking~\cite{vw86,vw94},
programs and properties are modeled by finite automata. Let $A$ be a
finite automaton that models a program and let $B$ be a finite
automaton that models a specification that the program should satisfy.
Correctness is defined by the language inclusion $\L(A) \subseteq \L(B)$,
that is all traces of the program (executions) should be traces of the
specification.  To solve the inclusion problem, the classical automata-theoretic 
solution constructs an automaton for $\L^c(B)$ the complement of
the language of the automaton $B$ and then checks
that $\L(A) \cap {\L^c(B)}$ is empty (the later intersection being
computed as a synchronised product).  

In the finite case, the program and the specification are
automata over finite words (NFA) and the construction for the
complementation is conceptually simple: it is achieved by a classical
subset construction.  In the case of infinite words, the program and
(or at least) the specification are nondeterministic B\"uchi automata
(NBW). The NBW are also complementable; this was first proved
by B\"uchi~\cite{buechi62}.  However, the result is
much harder to obtain than in the case of NFA.  The original
construction of B\"uchi has a $2^{O({2^n})}$ worst case complexity (where~$n$ 
is the size of the automaton to complement) which is not optimal.
In the late eighties Safra in~\cite{Safra88}, and later Kupferman and
Vardi in~\cite{kupferman97weak}, have given optimal complementation
procedures that have $2^{O({n \log n})}$ complexity (see~\cite{Michel88} for the
lower bound).  While for finite words, the classical algorithm
has been implemented and shown practically usable, for infinite
words, the theoretically optimal solution is difficult to
implement and very few results are known about their practical
behavior. 
%The actual attempts to implement them have shown very limited
%in the size of the specifications that can be handled: 
Recent implementations have shown that applying these algorithms for automata 
with more than around ten states is hard~\cite{TabakovV07,GurumurthyKSV03}.
Such sizes are clearly not sufficient in practice. 
As a consequence, tools like {\sc Spin}~\cite{spin} that
implement the automata-theoretic approach to model-checking ask
either that the complement of the specification is explicitly given
or they limit the specification to properties that are expressible in {\sc LTL}.

In this paper, we propose a new approach to check $\L(A) \subseteq \L(B)$ 
that can handle much larger B\"uchi automata. In a recent paper, 
we have shown that the classical subset construction can
be avoided and kept implicit for checking language inclusion and
language universality for NFA and their alternating extensions~\cite{cav06}. 
Here, we adapt and extend that technique to the more intricate case of
automata on infinite words.

To present the intuition behind our new techniques, let us consider a
simpler setting of the problem. Assume that we
are given a NBW $B$ and we want to check if $\Sigma^{\omega}
\subseteq \L(B)$, that is to check if $\L(B)$ is universal.  First,
remember that $\L(B)$ is universal when its complement $\L^c(B)$ is empty.  The
classical algorithm first complements $B$ and then checks for
emptiness. The language of a NBW is nonempty if there exists an
infinite run of the automaton that visits accepting locations
infinitely often.  The existence of such a run can be established
in polynomial time by computing the following fixed point $\F \equiv \nu y \cdot \mu x \cdot ( \Pre(x)
\cup ( \Pre(y) \cap \alpha ))$ where $\Pre$ is the predecessor
operator of the automaton (given a set $L$ of locations it returns the
set of locations that can reach $L$ in one step) and $\alpha$ is the set
of accepting locations of the automaton. The automaton is non-empty if
and only if its initial location is a member of the fixed
point $\F$.  This well-known algorithm is quadratic in the size of the automaton.  
Unfortunately, the automaton that accepts the language $\L^c(B)$ is usually 
huge and the evaluation of the fixed point is unfeasible for all but the smallest 
specifications~$B$.  To overcome this difficulty, we make the following observation:
if $\preceq$ is a \emph{simulation} pre-order on the locations of $B^c$ 
($\l_1 \preceq \l_2$ means $\l_1$ can simulate $\l_2$) which is compatible with
the accepting condition (if $\l_1 \preceq \l_2$ and $\l_2 \in \alpha$
then $\l_1 \in \alpha$), then the sets that are computed during the
evaluation of $\F$ are all \emph{$\preceq$-downward-closed} (if an element
$\l$ is in the set then all $\l' \preceq \l$ are also in the set). Then
$\preceq$-downward-closed sets can be represented by their $\preceq$-maximal elements 
and if operations on such sets can be computed directly on their representation, 
we have the ingredients to evaluate the fixed point in a more efficient way.
For an automaton~$\B$ over finite words, set inclusion would be a typical example 
of a simulation relation for~$\B^c$~\cite{cav06}. The same technique can be applied
to avoid subset constructions in games of imperfect information~\cite{DDR06,CDHR07}.
We generically call \emph{antichain algorithms} the techniques that are based
on compact representation of downward-closed because when the simulation
is a partial order (and it usually is), the maximal elements form an antichain,
i.e., a set of incomparable elements.

We show that the classical constructions for B\"uchi automata that are
used in the automata-theoretic approach to model-checking are all
equipped with a simulation pre-order that exists by construction and
does not need to be computed. On that basis we propose antichain algorithms
to check universality of NBW, language inclusion for NBW, and
emptiness of alternating B\"uchi automata (ABW). Each of these problems
reduces to emptiness checking of NBW, via classcial constructions.

The novelty of our antichain algorithms is to realize that only downward-closed
sets can be computed by the fixed point for emptiness, and therefore to use more 
succinct representations of those downward-closed sets, by storing maximal elements only. 
Moreover, such compact representations do not come at the price of an increase 
in the time complexity for the basic operations that are necessary to check emptiness
(such as $\cap$, $\cup$, and $\Pre$), i.e., we show that they are computable in time
polynomial in the size of the compact representation, while this size can be exponentially
smaller than the actual downward-closed set.
Note that, while a compact representation exists in general (i.e., for any
simulation pre-order), we have no generic result that would show that efficient
computations can be done symbolically in all cases. Therefore, we have to instantiate
the approach for each class of problem, and find efficient algorithms for
the basic operations.

% and a new algorithm to check satisfiability and to solve the model-checking problem of LTL.

We evaluate an implementation of our algorithm for the
universality problem of NBW and on a randomized model recently
proposed by Tabakov and Vardi.  We show that the performance of the
antichain algorithm on this randomized model outperforms by several order of
magnitude the existing implementations of the Kupferman-Vardi
algorithm~\cite{TabakovV07,GurumurthyKSV03}.  While the classical
solution is limited to automata of size 8 for some parameter values of
the randomized model, we are able to handle automata with more than
one hundred locations for the same parameter values. We have
identified the hardest instances of the randomized model for our
algorithms and show that we can still handle problems with several
dozens of locations for those instances.

\paragraph{{\it Structure of the paper.}}
In Section~\ref{sec:definitions}, we give all necessary definitions related to 
B\"uchi automata, and we recall the Kupferman-Vardi and
Miyano-Hayashi constructions that are used for complementation of NBW.
The reader interested in the general theory behind our technique can read 
Section~\ref{sec:fixed-point} without going into the details of those constructions
(only the definitions of NBW and emptiness are useful to understand Section~\ref{sec:fixed-point}). 
The notion of simulation pre-order for a B\"uchi automaton is presented and we prove 
that the fixed point needed to establish emptiness of nondeterministic B\"uchi automata
handles only downward closed sets for such pre-orders. 
We use this observation in Section~\ref{sec:emptiness-alternating} to define an antichain algorithm
to decide emptiness of ABW. In Section~\ref{sec:universality}, we
adapt the technique for the universality problem of NBW. In
Section~\ref{sec:implementation}, we report on the performances of the
algorithm for universality, and in Section~\ref{sec:inclusion-NBW},
we extend those ideas to obtain an antichain algorithm for language inclusion
of NBW. 

% Section 2

\section{B\"uchi Automata and Classical Algorithms}\label{sec:definitions}

\begin{defi}\label{def:ABW}
An \emph{alternating B\"uchi automaton} (ABW) is a tuple 
$\A = \tuple{\Loc, \iota, \Sigma, \delta, \alpha}$ where: 

\begin{enumerate}[$\bullet$]

\item $\Loc$ is a finite set of states (or locations). 
The \emph{size} of $\A$ is $\abs{\A} = \abs{\Loc}$; 

\item $\iota \in \Loc$ is the \emph{initial} state;

\item $\Sigma$ is a finite \emph{alphabet};

\item $\delta: \Loc \times \Sigma \to \B^+(\Loc)$ is the \emph{transition function} 
where $\B^+(\Loc)$ is the set of positive boolean formulas over $\Loc$, 
{\it i.e.} formulas built from elements in  $\Loc \cup \{\true, \false\}$ 
using the boolean connectives $\land$ and $\lor$; 

\item $\alpha \subseteq \Loc$ is the set of accepting states.

\end{enumerate}
\end{defi}

% Intuitively, the transition function combines nondeterministic choices (disjunction) 

% and universal choices (conjunction). For example, a transition $\delta(\l,\sigma) = (\l_1 \land \l_2)

% \lor (\l_3 \land \l_4)$ means that the automaton accepts a suffix $w^i$ of a word $w$ from state $\l$, if it accepts $w^{i+1}$

% from both $\l_1$ and $\l_2$ or from both $\l_3$ and $\l_4$.

We say that a set $X \subseteq \Loc$ \emph{satisfies} a
formula $\varphi \in \B^+(\Loc)$ (noted $X \models \varphi$) iff the
truth assignment that assigns $\true$ to the members of $X$ and
assigns $\false$ to the members of $\Loc \backslash X$ satisfies
$\varphi$.  A \emph{run} of $\A$ on an infinite word $w =
\sigma_0\cdot\sigma_1 \dots$ is a \DAG\/ $T_w = \tuple{V, v_{\iota},
  \to}$ where:

\begin{enumerate}[$\bullet$]

\item $V = \Loc \times \nat$ is the set of nodes. A node $(\l,i)$ represents the
state $\l$ after the first $i$ letters of the word $w$ have been read by $\A$. 
Nodes of the form $(\l,i)$ with $\l \in \alpha$ are called \emph{$\alpha$-nodes};

\item $v_{\iota} = (\iota,0) \in V$ is the root of the \DAG;

\item and $\to\, \subseteq V \times V$ is such that \begin{compressEnum} 
\itCompress if $(\l,i)\to(\l',i')$ then $i'=i+1$ and 
\itCompress for every $(\l,i) \in V$, the set $\{\l' \mid (\l,i) \to (\l',i+1)\}$ 
satisfies the formula $\delta(\l,\sigma_i)$.
\end{compressEnum}
We say that $(\l',i+1)$ is a \emph{successor} of $(\l,i)$ if $(\l,i) \to (\l',i+1)$, 
and we say that $(\l',i')$ is \emph{reachable} from $(\l,i)$ if $(\l,i) \to^* (\l',i')$.

\end{enumerate}
\noindent
A run $T_w = \tuple{V, v_{\iota}, \to}$ of $\A$ on an infinite word $w$ is
\emph{accepting} iff all its infinite paths $\pi$ rooted at $v_{\iota}$
%(thus $\pi \in \Loc^\omega$) 
visit $\alpha$-nodes infinitely often.
% are such that $\inf(\pi) \cap \alpha \neq \emptyset$ (B\"uchi condition). 
An infinite word $w \in \Sigma^\omega$ is \emph{accepted} by~$\A$ if
there exists an accepting run on it. We denote by~$\L(\A)$ the set
of infinite words accepted by~$\A$, and by~$\L^c(\A)$  % = \Sigma^{\omega} \setminus \L(\A)
the set of infinite words that are not accepted by~$\A$.

\begin{defi}\label{def:NBW}
A \emph{nondeterministic B\"uchi automaton} (NBW) is an ABW whose
transition function is restricted to disjunctions over $\Loc$. 
\end{defi}

Runs of NBW reduce to (linear) traces.  The transition function of NBW is
often seen as a function $[Q \times \Sigma \to 2^Q]$ and we write
$\delta(\l,\sigma) = \{\l_1, \dots, \l_n\}$ instead of
$\delta(\l,\sigma) = \l_1 \lor \l_2 \lor \dots \lor \l_n$. 
We note by $\Pre^{\A}_{\sigma}(L)$ the set of \emph{predecessors} by $\sigma$ 
of the set $L$: $\Pre^{\A}_{\sigma}(L)=\{\l \in \Loc \mid \exists \l' \in L:
\l' \in \delta(\l,\sigma) \}$. Let $\Pre^{\A}(L) = \{\l \in \Loc \mid \exists
\sigma \in \Sigma: \l \in \Pre^{\A}_{\sigma}(L) \}$.

\paragraph{{\bf Problems}} 
The \emph{emptiness problem} for NBW is to decide, given an NBW $\A$,
whether $\L(\A) = \emptyset$. This problem is solvable in
polynomial time. The symbolic approach through
fixed point computation is quadratic in the size of $\A$~\cite{EmmersonLei86}.
Other symbolic approaches have been proposed with better complexity bounds~\cite{BloemGS00,GentiliniPP03}, 
but the fixed point computation shows better performances in practice~\cite{RaviBS00}.

The \emph{universality problem} for NBW is to decide, given an NBW
$\A$ over the alphabet $\Sigma$ whether $\L(\A) = \Sigma^{\omega}$
where $\Sigma^{\omega}$ is the set of all infinite words on $\Sigma$.
This problem is {\sc PSpace}-complete~\cite{SVW87}.  The classical algorithm to
decide universality is to first complement the NBW and then to check
emptiness of the complement. The difficult step is the complementation
as it may cause an exponential blow-up in the size of the automaton.
There exist two types of construction, one is based on a
determinization of the automaton~\cite{Safra88} and the other uses ABW
as an intermediate step~\cite{kupferman97weak}. We review the second 
construction below.

The \emph{language inclusion problem} for NBW is to decide, given two
NBW $\A$ and $\B$, whether $\L(\A) \subseteq \L(\B)$. This
problem is central in model-checking and it is {\sc PSpace}-complete in the size of $\B$. 
The classical solution consists in checking the
emptiness of $\L(\A) \cap \L^c(\B)$, which again requires the
expensive complementation of $\B$.

The \emph{emptiness problem} for ABW is to decide, given an ABW $\A$,
whether $\L(\A) = \emptyset$. This problem is also {\sc
  PSpace}-complete and it can be solved using a translation from ABW
to NBW that preserves the language of the automaton~\cite{MiyanoH84}.
Again, this construction involves an exponential blow-up that makes
explicit implementations feasible only for automata limited to around
ten states. However, the emptiness problem for ABW is very important
in practice for LTL model-checking as there exist efficient polynomial
translations from LTL formulas to ABW~\cite{GastinOddoux}. The classical
construction is presented below.

\paragraph{{\bf Kupferman-Vardi construction}} 
Complementation of ABW is straightforward by dualizing the transition
function (by swapping $\land$ and $\lor$, and swapping $\true$ and
$\false$ in each formulas) and interpreting the accepting condition
$\alpha$ as a co-B\"uchi condition, {\it i.e.} a run $T_w$ is accepted
if all its infinite paths % $\pi$ rooted at the initial state of the automaton 
have a suffix that contains no $\alpha$-nodes.
%visit $\alpha$-nodes only finitely many times.

The result is an alternating co-B\"uchi automaton (ACW). The accepting runs of
ACW have a layered structure that has been studied
in~\cite{kupferman97weak}, where the notion of \emph{rank} is
defined.  The rank is a nonnegative integer associated to each node of an accepting
run $T_w$ of an ACW on a word $w$.  Let $G_0 = T_w$. Nodes of rank $0$
are those nodes from which only finitely many nodes are reachable in
$G_0$. Let $G_1$ be the run $T_w$ from which all nodes of rank $0$
have been removed. Then, nodes of rank $1$ are those nodes of $G_1$
from which no $\alpha$-node is reachable in $G_1$. For all $i\geq 2$, let
$G_{i}$ be the run $T_w$ from which all nodes of rank
$0,\dots,i-1$ have been removed. Then, nodes of rank $2i$ are those
nodes of $G_{2i}$ from which only finitely many nodes are reachable in
$G_{2i}$, and nodes of rank $2i+1$ are those nodes of $G_{2i+1}$ from
which no $\alpha$-node is reachable in $G_{2i+1}$. Intuitively, the
rank of a node $(\l,i)$ hints how difficult it is to prove that all
the paths of $T_w$ that start in $(\l,i)$ visit $\alpha$-nodes only
finitely many times. It can be shown that every node has a rank
between $0$ and $2(\abs{\Loc} - \abs{\alpha})$, and all $\alpha$-nodes
have an even rank~\cite{GurumurthyKSV03}. The layered structure of the 
runs of ACW induces a construction to complement ABW~\cite{kupferman97weak}. 
We present this construction directly for NBW. 

\begin{defi}[\cite{kupferman97weak}]\label{def:KV-construction}
Given a NBW $\A = \tuple{\Loc, \iota, \Sigma, \delta,
  \alpha}$ and an even number $k \in \nat$, let $\KV(\A,k) = \tuple{\Loc',
  \iota', \Sigma, \delta', \alpha'}$ be an ABW such that:

\begin{enumerate}[$\bullet$]
\item $\Loc' = \Loc \times [k]$ where $[k] = \{0,1,\dots,k\}$.
  Intuitively, the automaton $\KV(\A,k)$ is in state $(\l,n)$
  after the first $i$ letters of the input word $w$ have been read if
  it guesses that the rank of the node $(\l,i)$ in a run of $\A$ on
  $w$ is at most $n$;
\item $\iota'  = (\iota, k)$;
\item $\delta'((\l,i),\sigma) = \left\{\begin{array}{ll} \false  & \text{ if } \l\in\alpha \text{ and } i \text{ is odd} \\ 
						\bigwedge_{\l' \in \delta(\l,\sigma)} \bigvee_{0 \leq i'\leq i} (\l',i') & \text{ otherwise }
				\end{array}\right.$ 

% $\delta'((\l,i),\sigma) = \false$ if $\l\in\alpha$ and $i$ is odd, and otherwise 
% $\delta'((\l,i),\sigma) = \bigvee_{i'\leq i} (\l_1,i') \land \bigvee_{i'\leq i} (\l_2,i') \land \dots \land \bigvee_{i'\leq i} (\l_n,i')$
% if $\delta(\l,\sigma) = \l_1 \lor \l_2 \lor \dots \lor \l_n$;
\medskip
\noindent For example, if $\delta(\l,\sigma) = \{\l_1, \l_2\}$, then 
$$\delta'((\l,2),\sigma) = ((\l_1,2) \lor (\l_1,1) \lor (\l_1,0)) \land ((\l_2,2) \lor (\l_2,1) \lor (\l_2,0))$$
\item $\alpha' = \Loc \times [k]^{odd}$ where $[k]^{odd}$ is the set of odd numbers in $[k]$.
\end{enumerate}
\end{defi}

The ABW specified by the Kupferman-Vardi construction accepts the
complement language of $\L(\A)$ and its size is quadratic in the size of the
original automaton~$\A$.

\begin{thm}[\cite{kupferman97weak}]\label{theo:complement}
For all NBW $\A = \tuple{\Loc, \iota, \Sigma, \delta,
\alpha}$, for all $0\leq k' \leq k$, we have $\L(\KV(\A,k')) \subseteq \L(\KV(\A,k))$ and 
for $k = 2(\abs{\Loc} - \abs{\alpha})$, we have $\L(\KV(\A,k)) = \L^c(\A)$.
\end{thm}

\paragraph{{\bf Miyano-Hayashi construction}}
Classically, to check emptiness of ABW, a variant 
of the subset construction is applied that transforms the ABW into a NBW 
that accepts the same language~\cite{MiyanoH84}. Intuitively, the NBW 
maintains a set $s$ of states of the ABW that corresponds to a whole
level of a guessed run \DAG\/ of the ABW. In addition, the NBW maintains
a set $o$ of states that ``owe'' a visit to an accepting state. Whenever
the set $o$ gets empty, meaning that every path of the guessed run has visited
at least one accepting state, the set $o$ is initiated with the current
level of the guessed run. It is asked that $o$ gets empty infinitely often
in order to ensure that every path of the run \DAG\/ visits accepting states
infinitely often. The construction is as follows.

\begin{defi}[\cite{MiyanoH84}]\label{def:MH-construction}
Given an ABW $\A = \tuple{\Loc, \iota, \Sigma, \delta,
\alpha}$, define $\MH(\A)$ as the NBW $\tuple{2^{\Loc} \times 2^{\Loc},
(\{\iota\},\emptyset), \Sigma, \delta', \alpha'}$ where
$\alpha' = 2^{\Loc} \times \{\emptyset\}$ and $\delta'$ is defined,
for all $\tuple{s,o} \in 2^{\Loc} \times 2^{\Loc}$ and $\sigma \in
\Sigma$, as follows:

\begin{enumerate}[$\bullet$] 

\item If $o\neq \emptyset$, then
$$\delta'(\tuple{s,o},\sigma) = \{ \tuple{s',o'\setminus \alpha} \mid o'\subseteq s', 
s' \models \bigwedge_{\l \in s} \delta(\l,\sigma) \textrm{ and }\penalty-1000 o' \models \bigwedge_{\l \in o}
\delta(\l,\sigma)\}$$

\item If $o= \emptyset$, then $\delta'(\tuple{s,o},\sigma) = \{ \tuple{s',s'\setminus \alpha} \mid  
s' \models \bigwedge_{\l \in s} \delta(\l,\sigma)\}$.
\end{enumerate} 

\end{defi}

The size of the Miyano-Hayashi construction is exponential in the 
size of the original automaton. 

\begin{thm}[\cite{MiyanoH84}]\label{theo:subset-construction}
For all ABW $\A$, we have $\L(\MH(\A)) = \L(\A)$.
\end{thm}

The size of the automaton obtained after the Kupferman-Vardi
and the Miyano-Hayashi construction is an obstacle to the
direct implementation of the method.  
%In Section~\ref{sec:fixed-point}, we
%propose a new approach that circumvents this problem.

\paragraph{{\bf Direct complementation}}\label{sec:direct-complementation}
In our solution, we \emph{implicitly} use the two constructions to
complement B\"uchi automata but, as we will see, we do not construct
the automata. For the sake of clarity, we give below the
specification of the automaton that would result from the composition of
the two constructions. In the definition of the state space, we omit the
states $(\l,i)$ for $\l \in \alpha$ and $i$ odd, as those states have no 
successor in the Kupferman-Vardi construction.
 
% We rephrase the subset construction from~\cite{MiyanoH84} directly in terms of 
% the NBW $\A$ that we complement.  

\begin{defi}\label{def:direct-complementation}
Given a NBW $\A = \tuple{\Loc, \iota, \Sigma, \delta, \alpha}$ 
and an even number $k \in \nat$, let 
$\KVMH(\A,k) = \tuple{Q_k \times Q_k, q_{\iota}, \Sigma, \delta', \alpha'}$ be a NBW such that:
\begin{enumerate}[$\bullet$] 
\item $Q_k = 2^{(\Loc \times [k]) \setminus (\alpha \times \nat^{odd})}$ where
$\nat^{odd}$ is the set of odd natural numbers;
\item $q_{\iota} = (\{(\iota, k)\}, \emptyset)$;
% \item Let $\odd = \Loc \times [k]^{odd}$;
\item Let $\odd = \Loc \times [k]^{odd}$; $\delta'$ is defined for all $s,o \in Q_k$ and 
$\sigma \in \Sigma$, as follows:
\begin{enumerate}[$-$] 
\item If $o \neq \emptyset$, then $\delta'(\tuple{s,o},\sigma)$ is the set of pairs $\tuple{s',o'\setminus \odd}$ such that:
	\begin{enumerate}[$(i)$] 
	\item $o'\subseteq s'$; 
	\item $\forall (\l,n) \in s \cdot \forall \l' \in \delta(\l,\sigma)\cdot \exists n' \leq n: (\l',n') \in s'$;
%		\begin{itemize}
%		\item[$(a)$] $\forall \l' \in \delta(\l,\sigma)\cdot \exists (\l',n') \in s': n' \leq n$ and
%		\item[$(b)$] $\l \not\in \alpha \lor n$ is even 
%		\end{itemize}
	\item $\forall (\l,n) \in o \cdot \forall \l' \in \delta(\l,\sigma)\cdot \exists n' \leq n: (\l',n') \in o'$.
%		\begin{itemize}
%		\item[$(a)$] $\forall \l' \in \delta(\l,\sigma)\cdot \exists (\l',n') \in o': n' \leq n$ and
%		\item[$(b)$] $\l \not\in \alpha \lor n$ is even $\}$
%		\end{itemize}
	\end{enumerate}
\item If $o= \emptyset$, then $\delta'(\tuple{s,o},\sigma)$ is the set of pairs $\tuple{s',s'\setminus \odd}$ such that:
	\begin{enumerate}[$(i)$]
	\item[] $\forall (\l,n) \in s \cdot \forall \l' \in \delta(\l,\sigma)\cdot \exists n' \leq n: (\l',n') \in s'$.
	\end{enumerate}
\end{enumerate} 
\item $\alpha' = Q_k \times \{\emptyset\}$;
\end{enumerate} 
\end{defi}

 We write $\tuple{s,o} \xrightarrow{\sigma}_{\delta'} \tuple{s',o'}$ to 
denote $\tuple{s',o'} \in \delta'(\tuple{s,o},\sigma)$.

\begin{thm}[\cite{kupferman97weak,MiyanoH84}]\label{theo:direct-complement}
For every NBW $\A = \tuple{\Loc, \iota, \Sigma, \delta, \alpha}$ and
for all $\,0\leq k' \leq k$, we have $\L(\KVMH(\A,k')) \subseteq
\L(\KVMH(\A,k))$.  In case of $k = 2(\abs{\Loc} - \abs{\alpha})$, we
also have $\L(\KVMH(\A,k)) = \L^c(\A)$.
\end{thm}

In the sequel, we denote by $\KVMH(\A)$ the automaton $\KVMH(\A,2(\abs{\Loc} - \abs{\alpha}))$,
and we denote by $Q \times Q$ its set of states (we omit the subscript $k$).

% Section 3
\section{Simulation Pre-Orders and Fixed Points}\label{sec:fixed-point}

Let $\A = \tuple{\Loc, \iota, \Sigma, \delta, \alpha}$ be a NBW.  Let
$\langle 2^{\Loc},\subseteq, \cup,\cap,\emptyset,\Loc \rangle$ be the
powerset lattice of locations. The fixed point formula
% \begin{center}
  $\F_{\A} \equiv \nu y \cdot \mu x \cdot ( \Pre^{\A}(x) \cup ( \Pre^{\A}(y) \cap \alpha ))$
% \end{center}\mbox{}\\[-16pt]
% \noindent
can be used to check emptiness of $\A$ as we have $\L(\A)\not= \emptyset$ iff $\iota \in \F_{\A}$.
Intuitively, the greatest fixed point $\nu y$ in $\F_{\A}$ computes in the
$n$-th iteration the set of states from which $n$ accepting states can be visited
with some word. When this set stabilizes, infinitely many visits to an accepting
state are possible.
% \begin{thm}\cite{EmmersonLei86,Dam94}\label{theo:buchi-fixed-point}
%   For all NBW $\A$ with initial location $\iota$, we
%   have $\L(\A)\not= \emptyset$ iff $\iota \in \F_{\A}$.
% \end{thm}

We show in this section that a certain structural property of the NBW
is tightly correlated to the structure of the sets that are computed by the 
fixed point $\F_{\A}$. The key property is the notion of simulation relation
for finite automata. Let $\preceq \subseteq \Loc \times \Loc$ be a pre-order and let $\l_1
\prec \l_2$ iff $\l_1 \preceq \l_2$ and $\l_2 \not\preceq \l_1$.

\begin{figure}[!tb]
  \unitlength=.8mm
\def\fsize{\normalsize}

\hrule
\begin{picture}(160,46)(0,0)
%\put(0,-3){\framebox(160,48){}}

{\fsize

\node[Nmarks=n](q1)(20,38){$\l_1$}
\node[Nmarks=n](q2)(60,38){$\l_2$}
\node[Nmarks=n](q3)(20,8){$\l_3$}
\node[Nframe=n](dummy)(5,23){If}
\node[Nframe=n, Nadjust=wh, Nadjustdist=2.2](dummy)(20,23){\rotatebox{90}{$\preceq$}}
%\node[Nmarks=r](q4)(50,10){$\l_4$}

\node[Nmarks=n](r2)(150,38){$\l_2$}
\node[Nmarks=n](r3)(110,8){$\l_3$}
\node[Nmarks=n](r4)(150,8){$\l_4$}
\node[Nframe=n](eummy)(83,23){then}
\node[Nframe=n, Nadjust=wh, Nadjustdist=2.2](eummy)(150,23){\rotatebox{90}{$\preceq$}}

\drawedge[ELpos=50, ELside=l, ELdist=1, curvedepth=0](q1,q2){$\sigma$}
\drawedge[ELpos=50, ELside=l, ELdist=1, curvedepth=0](r3,r4){$\sigma$}

\drawedge[dash={1}0, ELpos=50, ELside=l, ELdist=1, curvedepth=0, AHnb=0](q1,dummy){}
\drawedge[dash={1}0, ELpos=50, ELside=l, ELdist=1, curvedepth=0, AHnb=0](q3,dummy){}

\drawedge[dash={1}0, ELpos=50, ELside=l, ELdist=1, curvedepth=0, AHnb=0](r2,eummy){}
\drawedge[dash={1}0, ELpos=50, ELside=l, ELdist=1, curvedepth=0, AHnb=0](r4,eummy){}

%\drawloop[ELside=l,loopCW=y, loopdiam=6](q0){$b$}
%\drawloop[ELside=l,loopCW=y, loopdiam=6](q1){$b$}
%\drawloop[ELside=l,loopCW=y, loopdiam=6](q2){$b$}
%\drawloop[ELside=l,loopCW=y, loopdiam=6](q3){$b$}
%\drawloop[ELside=l,loopCW=y, loopdiam=6](q4){$b$}
%\drawloop[ELside=l,loopCW=y, loopangle=0, loopdiam=6](q6){$a,b$}

%\drawedge[ELpos=55, ELside=r, ELdist=1, curvedepth=-18](q3,q0){$a,b$}
%\drawbpedge[ELpos=78, ELside=r, ELdist=1, syo=3, exo=-1](q3,165,20,q0,80,50){$a,b$}

%\drawedge[dash={1}0](n3bis,nkbis){$0,1$}
}
\end{picture}
\hrule
  %\caption{\protect\parbox[t]{40mm}{Normalization widget.}} \label{fig:widgetNormalizing}
  %\caption{Alternating automaton for $\varphi \equiv \lnot (GF p \to G(\lnot p \to Fr))$.}
  %\caption{$\A_{\varphi}$.}
  \caption{Simulation (Definition~\ref{def:simulation}).}
  \label{fig:simulation}
\end{figure}
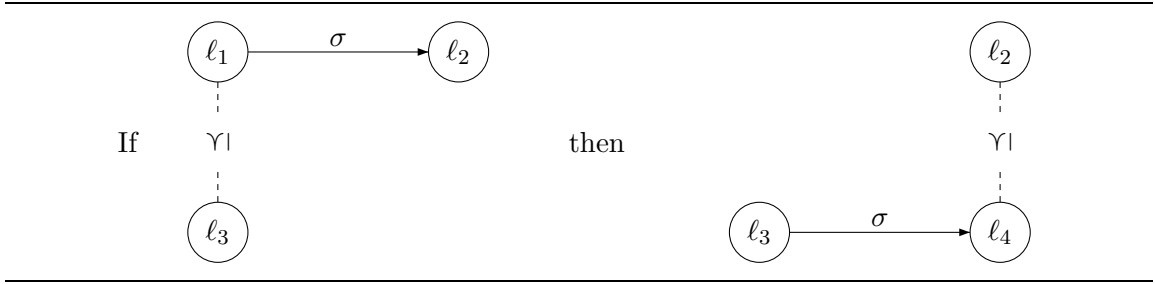

\begin{defi}\label{def:simulation}
  A pre-order $\preceq$ is a \emph{simulation} for $\A$ iff the following
  properties hold:
  \begin{enumerate}[$\bullet$]
    
    \item for all $\l_1,\l_2,\l_3 \in \Loc$, for all $\sigma \in \Sigma$, if
    $\l_3 \preceq \l_1$ and $\l_2 \in \delta(\l_1,\sigma)$ then there exists
    $\l_4 \in \Loc$ such that $\l_4 \preceq \l_2$ and $\l_4 \in \delta(\l_3,\sigma)$ (see illustration in \figurename~\ref{fig:simulation});

    \item for all $\l \in \alpha$, for all $\l' \in \Loc$, if $\l' \preceq \l$ 
     then $\l' \in \alpha$.

  \end{enumerate}
\end{defi}

\paragraph{{\bf Downward-closed sets}} A set $L \subseteq \Loc$ is \emph{$\preceq$-closed} iff for all
$\l_1,\l_2 \in \Loc$, if $\l_1 \preceq \l_2$ and $\l_2 \in L$ then $\l_1 \in
L$. The \emph{$\preceq$-closure} of $L$, is the set $\closure{L}=\{ \l
\in \Loc \mid \exists \l' \in L : \l \preceq \l' \}$.  We denote by 
$\Max(L)$ the set of \emph{$\preceq$-maximal} elements of $L$: 
$\Max(L)=\{ \l \in L \mid \nexists \l' \in L : \l \prec \l' \}$. 
% When the context is ambiguous, we sometimes write $\closure_{\preceq}$
% and $\Max_{\preceq}$ with the intended pre-order in subscript.
For any $\preceq$-closed set $L \subseteq \Loc$, we have 
$L=\closure{\Max(L)}$. Furthermore, if $\preceq$ is a partial order, 
then $\Max(L)$ is an \emph{antichain} of elements and it can serve as a canonical 
representation of~$L$. 

Our goal is to show that the operators involved in the fixed
point formula $\F_{\A}$ preserve $\preceq$-closedness. This is
true for union and intersection, for all relations $\preceq$.

\begin{lem}\label{lem:closure-properties-a}
  For all relations $\preceq$, for all $\preceq$-closed sets $L_1,L_2$,
  the sets $L_1 \cup L_2$ and $L_1 \cap L_2$ are $\preceq$-closed.
\end{lem}
 
The next lemma shows that simulation relations are necessary (and also sufficient) to
guarantee preservation of $\preceq$-closedness under the $\Pre$ operator.
Note that many other notions of simulation pre-orders have been defined 
for B\"uchi automata, see~\cite{simulations}.\footnote{In ~\cite{simulations},
the simulation of Definition~\ref{def:simulation} is called \emph{direct simulation}.}

\begin{lem}\label{lem:closure-properties-b}
  Let $\A = \tuple{\Loc, \iota, \Sigma, \delta, \alpha}$ be a NBW.
  A pre-order $\preceq \subseteq \Loc \times \Loc$ is a simulation for $\A$
  if and only if the following two properties hold:
  \begin{enumerate}[\em(a)]
    \item the set $\alpha$ is $\preceq$-closed.
    \item for all $\preceq$-closed sets $L \subseteq \Loc$, for all $\sigma \in \Sigma$, $\Pre^{\A}_{\sigma}(L)$ is $\preceq$-closed;
  \end{enumerate}
\end{lem}

\proof 
First, assume that $\preceq$ is a simulation for $\A$.
Then, the set $\alpha$ is $\preceq$-closed by Definition~\ref{def:simulation}, which establishes (a).
To prove (b), let $L \subseteq \Loc$ be a $\preceq$-closed set and let $\sigma \in \Sigma$.
For all $\l_1 \in \Pre^{\A}_{\sigma}(L)$ there exists $\l_2 \in L$ such that 
$\l_2 \in \delta(\l_1,\sigma)$. By Definition~\ref{def:simulation},
for all $\l_3 \preceq \l_1$ there exists $\l_4 \in \Loc$ such that 
$\l_4 \preceq \l_2$ and $\l_4 \in \delta(\l_3,\sigma)$ (see \figurename~\ref{fig:simulation}). 
So $\l_4 \in L$ since $L$ is $\preceq$-closed and $\l_2 \in L$, and thus $\l_3 \in \Pre^{\A}_{\sigma}(L)$
which shows that $\Pre^{\A}_{\sigma}(L)$ is $\preceq$-closed. 

Second, assume that (a) and (b) hold, and show that $\preceq$ satisfies Definition~\ref{def:simulation}.
By (a), for all $\l \in \alpha$ and for all $\l' \preceq \l$, we have $\l' \in \alpha$.
Now, let $\l_1,\l_2,\l_3 \in \Loc$ and $\sigma \in \Sigma$ such that 
$\l_3 \preceq \l_1$ and $\l_2 \in \delta(\l_1,\sigma)$. Consider the 
$\preceq$-closed set $L_2 = \closure{\{\l_2\}}$.
By (b), the set $\Pre^{\A}_{\sigma}(L_2)$ is $\preceq$-closed and thus
$\l_3 \in \Pre^{\A}_{\sigma}(L_2)$. Therefore, there exists $\l_4 \in L_2$ (i.e. 
$\l_4 \preceq \l_2$) such that $\l_4 \in \delta(\l_3,\sigma)$. 
Hence, $\preceq$ is a simulation for $\A$.
\qed

%As a direct consequence of the previous lemma, we have:

%\begin{thm}
%  For all nondeterministic B\"uchi automata $\A = \tuple{\Loc, \l_0, \Sigma, \delta,
%    \alpha}$, for all adequate pre-orders $\preceq$, the set $\F_{\A}$
%  defined by $\nu y \cdot \mu x \cdot ( \Pre^{\A}(x) \cup (
%  \Pre^{\A}(y) \cap \alpha ))$ is $\preceq$-closed.

%\end{thm}

\begin{figure}[!tb]
  \unitlength=.8mm
\def\fsize{\normalsize}

\hrule
\begin{picture}(115,85)(0,0)
%\put(0,-3){\framebox(115,90){}}

{\fsize

\gasset{Nw=9,Nh=9,Nmr=4.5}

\node[Nmarks=n](q1)(85,75){$\l$}

\node[Nmarks=n](q2)(80,60){}%{2}
\node[Nmarks=n](q3)(93,50){}%{3}
\node[Nmarks=n](q4)(71,38){}%{4}
\node[Nmarks=n](q5)(100,32){}%{5}
\node[Nmarks=n](q6)(88,25){}%{6}
\node[Nmarks=n](q7)(66,15){}%{7}
\node[Nmarks=n](q8)(95,8){}%{8}

\node[Nframe=n](left)(55,-1){}
\node[Nframe=n](right)(115,-1){}
\drawbpedge[dash={1.5 2.5}0, AHnb=0](left,82,112,right,98,112){}

\node[Nmarks=n](l1)(10,48){$\l_1$}
\node[Nmarks=n](l2)(40,40){$\l_2$}
\node[Nframe=n](dots)(25,32){$\vdots$}
\node[Nmarks=n](l3)(25,18){$\l_k$}

\drawedge[ELpos=25, ELside=l, ELdist=1, curvedepth=10](l1,q5){$\sigma$}
\drawedge[ELpos=50, ELside=l, ELdist=1, curvedepth=6](l2,q2){$\sigma$}
\drawedge[ELpos=50, ELside=l, ELdist=1, curvedepth=6](l3,q7){$\sigma$}

%\drawedge[ELpos=50, ELside=l, ELdist=1, curvedepth=0](r3,r4){$\sigma$}

%\drawedge[dash={1}0, ELpos=50, ELside=l, ELdist=1, curvedepth=0, AHnb=0](q1,dummy){}

%\drawloop[ELside=l,loopCW=y, loopdiam=6](q0){$b$}
%\drawloop[ELside=l,loopCW=y, loopdiam=6](q1){$b$}
%\drawloop[ELside=l,loopCW=y, loopdiam=6](q2){$b$}
%\drawloop[ELside=l,loopCW=y, loopdiam=6](q3){$b$}
%\drawloop[ELside=l,loopCW=y, loopdiam=6](q4){$b$}
%\drawloop[ELside=l,loopCW=y, loopangle=0, loopdiam=6](q6){$a,b$}

%\drawedge[ELpos=55, ELside=r, ELdist=1, curvedepth=-18](q3,q0){$a,b$}
%\drawbpedge[ELpos=78, ELside=r, ELdist=1, syo=3, exo=-1](q3,165,20,q0,80,50){$a,b$}

%\drawedge[dash={1}0](n3bis,nkbis){$0,1$}
}
\end{picture}
\hrule
  %\caption{\protect\parbox[t]{40mm}{Normalization widget.}} \label{fig:widgetNormalizing}
  %\caption{Alternating automaton for $\varphi \equiv \lnot (GF p \to G(\lnot p \to Fr))$.}
  %\caption{$\A_{\varphi}$.}
  \caption{Computing the predecessors of a $\preceq$-closed set.}
  \label{fig:example-max}
\end{figure}
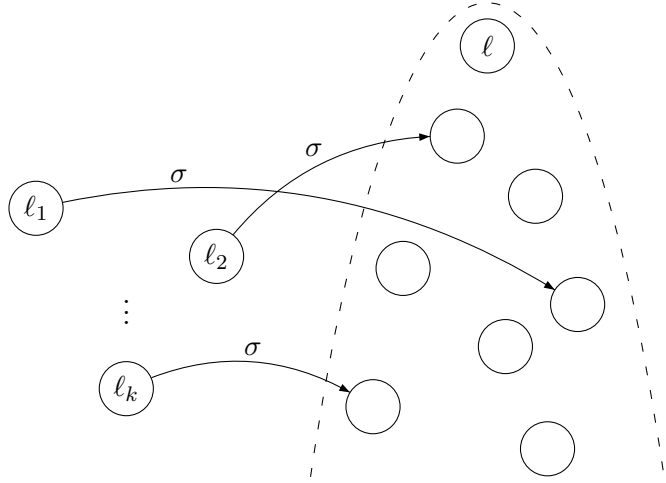

Lemmas~\ref{lem:closure-properties-a} and~\ref{lem:closure-properties-b} 
entail that all sets computed in the iterations of the fixed point formula~$\F_{\A}$
are $\preceq$-closed for any simulation $\preceq$ for $\A$.
We can take advantage of this fact to use a compact representation of those
sets, namely their maximal elements. This would indeed reduce the size of the 
sets to manipulate by the fixed point algorithms (possibly exponentially as we 
will see later). Notice that in general, this compact representation can make 
more difficult the computation of the $\Pre$ operator. To illustrate this, consider
the example in \figurename~\ref{fig:example-max} where we want to compute 
$\Pre_\sigma(\closure{\{\l\}})$. More precisely, given $\l$ we need to compute
the maximal elements of the $\preceq$-closed set $\Pre_\sigma(\closure{\{\l\}})$.
The set $\closure{\{\l\}}$ is delimited by the dashed curve in the figure. 
First, note that applying $\Pre_\sigma$ to $\{\l\}$ would give the empty set
from which the correct result can obviously not be extracted. Second, if we assume
that the states $\l_1, \dots, \l_k$ are $\preceq$-incomparable, then the result 
is $\Max(\Pre_\sigma(\closure{\{\l\}})) = \{ \l_1, \dots, \l_k \}$, which shows 
that essentially any set can be obtained, including sets of maximal elements that 
are huge or difficult to manipulate symbolically. Third, even if the result is compact 
(e.g., if $\l_i \preceq \l_1$
for all $1 \leq i \leq k$, then the result is the singleton $\{\l_1\}$), the computation
may somehow require to enumerate all the $\l_i$ for $i=1,2,\dots,k$ where $k$
may be for instance exponential in the size of the problem. 

The above remarks show that for \emph{each particular} application (i.e., for
each class of automata, and each particular simulation $\preceq$ that we use),
we need $(1)$ to define a predecessor operator $\Pre^{{\sf abs}}$
that applies to maximal elements, such that $\Pre^{{\sf abs}}(\Max(L)) = \Max(\Pre(L))$ 
for all $\preceq$-closed sets $L$, $(2)$ to present an algorithm to compute this 
operator, and establish its correctness, and $(3)$ to study the complexity of 
such an algorithm.

Finally, note that the way to compute $\Max(L_1 \cap L_2)$ given $\Max(L_1)$ 
and $\Max(L_2)$ should also be defined for each application, while for union,
the following general rule applies:
$\Max(L_1 \cup L_2) = \Max(\Max(L_1) \cup \Max(L_2))$.

In the next sections, we show that the NBW that we have to analyze
in the automata-based approach to model-checking are all equipped with
a simulation pre-order that can be exploited to compute efficiently
the intersection and the predecessor operators. Hence, we show that 
the expected efficiency in terms of space consumption 
of the antichain representation does not come at the price of a blow-up in the 
computation times of these operators. 
We do so for the emptiness problem of ABW,
and for the universality and language inclusion problems for NBW. All these problems
can be reduced to the emptiness problem of NBW that are obtained by specific 
constructions (analogous of the powerset construction), for which simulation relations 
\emph{need not to be computed} for each instance of the problems, but can be defined \emph{generically}
(like set inclusion is such a relation for the classical powerset construction).

% Intuitively, when computing the sequence of approximations for
% $\F_{\A}$, we can concentrate on maximal elements for a simulation
% pre-order as those locations are such that if they have an accepting
% run in $\A$, then all the locations that are smaller for the pre-order 
% also have an accepting run in $\A$.

% This lemma and this theorem are useful when, given an adequate
% pre-order $\preceq$, we are able to compute ${\sf Pre}$, $\cup$ and $\cap$ 
% operations by manipulating maximal elements only and when the
% set of maximal elements is (much) smaller than the entire set. We will
% see in the rest of the paper that the Buchi automata that we have to
% analyze in order to check universality or inclusion of omega-regular
% languages are all equipped with adequate pre-orders that can be
% exploited to manipulate maximal elements only during the evaluation of
% the fixed point expression ${\cal F}$ and never construct the entire
% automaton.

% Section 4

\section{Emptiness of ABW}\label{sec:emptiness-alternating}

We now show how to apply Lemmas~\ref{lem:closure-properties-a} and~\ref{lem:closure-properties-b} to check more efficiently the
emptiness of ABW. Let $\A_1=\tuple{\Loc_1, \iota_1, \Sigma, \delta_1,
  \alpha_1}$ be an ABW for which we want to decide whether
$\L(\A_1)=\emptyset$.  We know that the (exponential) Miyano-Hayashi
construction gives a NBW $\A_2=\MH(\A_1)$ such that $\L(\A_2) =
\L(\A_1)$. The emptiness of $\A_1$ (or equivalently of
$\A_2$) can be decided more efficiently by computing the fixed point
$\F_{\A_2}$ and without constructing explicitly $\A_2$. To do
so, we establish the existence of a simulation for $\A_2$ for which we can
compute $\cup$, $\cap$ and $\Pre$ by manipulating only maximal
elements of closed sets of locations.

% To answer this question efficiently, we show that the automaton
% $\A_2=\tuple{\Loc_1, \iota_1, \Sigma, \delta_1, \alpha_1}$ that is
% obtained from $\A_1$ using the Miyano-Hayashi construction, that is
% $\A_2={\sf MH}(\A_1)$, is such that there exists a simulation for $\A_2$
% such that the emptiness of $\A_2$ can be decided by computing
% $\F_{\A_2}$ without constructing explicitly $\A_2$ and by
% manipulating only maximal elements of closed sets.

\begin{defi}\label{def:simulation-for-MH}
Let $\MH(\A_1) = \tuple{\Loc_2, \iota_2, \Sigma, \delta_2, \alpha_2}$.
Remember that $\Loc_2 \subseteq 2^{\Loc_1} \times 2^{\Loc_1}$. 
Define the pre-order $\preceq_{{\sf alt}} \subseteq \Loc_2 \times
\Loc_2$ such that $\tuple{s,o} \preceq_{{\sf alt}} \tuple{s',o'}$ iff 
$(i)$ $s \subseteq s'$, $(ii)$ $o \subseteq o'$, and $(iii)$ $o=\emptyset$ iff $o'=\emptyset$. 
\end{defi}

Note that the pre-order $\preceq_{{\sf alt}}$ is a partial order. 
As a consequence, given a set of pairs $L=\{ \tuple{s_1,o_1},\tuple{s_2,o_2}, \dots, \tuple{s_n,o_n} \}$, 
the set $\Max(L)$ is an \emph{antichain} and identifies~$L$.

\begin{lem}\label{lem:alt-simulation-relation}
For all ABW $\A_1$, the partial order $\preceq_{{\sf alt}}$ is a simulation 
for $\MH(\A_1)$.
\end{lem}

\proof 
 % Let us show that $\preceq_{{\sf alt}}$ is a simulation for $\A_2$. 
  Let $\A_1=\tuple{\Loc_1, \iota_1, \Sigma, \delta_1, \alpha_1}$ and
  $\MH(\A_1) = \tuple{\Loc_2, \iota_2, \Sigma, \delta_2, \alpha_2}$.
  First, let $\sigma \in \Sigma$ and $\tuple{s_1,o_1}$, $\tuple{s_2,o_2}$, 
 $\tuple{s_3,o_3} \in \Loc_2$ be such that 
 $\tuple{s_1,o_1} \xrightarrow{\sigma}_{\delta_2} \tuple{s_2,o_2}$ and
 $\tuple{s_3,o_3} \preceq_{{\sf alt}} \tuple{s_1,o_1}$.
 We show that there exists $\tuple{s_4,o_4} \in \Loc_2$ such that 
 $\tuple{s_3,o_3} \xrightarrow{\sigma}_{\delta_2} \tuple{s_4,o_4}$
 and $\tuple{s_4,o_4} \preceq_{{\sf alt}} \tuple{s_2,o_2}$. 
 Let us consider the case where $o_1=\emptyset$. Then we have 
  $o_3=\emptyset$ by definition of $\preceq_{{\sf alt}}$ and 
  $\delta_2(\tuple{s_1,o_1},\sigma)=\{ \tuple{s',s' \setminus \alpha_1} \mid 
  s' \models \bigwedge_{l \in s_1} \delta_1(l,\sigma)\}$, 
  this set being contained in $\delta_2(\tuple{s_3,o_3},\sigma)=
  \{ \tuple{s',s' \setminus \alpha_1} \mid s' \models \bigwedge_{l \in s_3}
  \delta_1(l,\sigma) \}$ as $s_3$ puts less constraints than $s_1$ since
  $s_3 \subseteq s_1$. A similar reasoning holds if $o_1 \not=
  \emptyset$. Second, let $\tuple{s_1,o_1} \in \alpha_2$ and let $\tuple{s_2,o_2}
  \preceq_{{\sf alt}} \tuple{s_1,o_1}$. By definition of $\alpha_2$, we know
  that $o_1=\emptyset$, and by definition of $\preceq_{{\sf alt}}$ we have 
  $o_2=\emptyset$ and so $\tuple{s_2,o_2} \in \alpha_2$.
\qed

According to Lemmas~\ref{lem:closure-properties-a} and~\ref{lem:closure-properties-b}, 
all the sets that we compute to evaluate $\F_{\A_2}$ are \mbox{$\preceq_{{\sf alt}}$-closed}.  
We need to compute intersection and $\Pre$ by only manipulating maximal elements. 
Given $\tuple{s_1,o_1}, \tuple{s_2,o_2}$, we take $\tuple{s,o}$ such
that $\closure{\tuple{s,o}}=\closure{\tuple{s_1,o_1}}\, \cap \closure{\tuple{s_2,o_2}}$ as follows:
\begin{equation}\tuple{s,o} = \left\{\begin{array}{ll}\tuple{s_1 \cap s_2,o_1 \cap o_2} & \text{ if } o_1 \cap o_2 \not=\emptyset, \\
					\tuple{s_1 \cap s_2,\emptyset}    & \text{ if } o_1=o_2=\emptyset, \\
\end{array}\right.\label{eq:intersection}\end{equation}
and otherwise the intersection is empty. 

Algorithm~1 computes the maximal elements of the set of
$\sigma$-predecessors of the $\preceq_{{\sf alt}}$-clo\-sure of a pair
$\tuple{s',o'}$. This allows to compute the maximal elements of the
set of predecessors of any $\preceq_{{\sf alt}}$-closed set by just
manipulating its maximal elements, since $\Pre^{\A}(L_1 \cup L_2) =
\bigcup_{\sigma \in \Sigma} \Pre^{\A}_{\sigma}(L_1) \cup
\Pre^{\A}_{\sigma}(L_2)$.

Note that our algorithm runs in polynomial time, more precisely in $O(\abs{\Loc_1}\cdot\norm{\delta_1})$ 
where $\norm{\delta_1}$ is the size of the transition relation, defined as
the maximal number of boolean connectives in a formula $\delta_1(\l,\sigma)$.

\begin{algorithm}[!tbp]
  % {\scriptsize \SetKwComment{\tcc}{//}{}
  \AlgData{An ABW $\A_1 = \tuple{\Loc_1, \iota_1, \Sigma, \delta_1, \alpha_1}$, 
  $\sigma \in \Sigma$ and $\tuple{s',o'} \in 2^{\Loc_1} \times 2^{\Loc_1}$ such that $o' \subseteq s'$.}  

  \AlgResult{The $\preceq_{{\sf alt}}$-antichain $\Pre^{{\sf alt}}_{\sigma}(\tuple{s',o'})$.}
  \flushleft
  \Begin { 
    \nl $L_{\sf Pre} \gets \emptyset$\;
    % \nl \ForEach{$\sigma \in \Sigma$}
    % {
    %  \nl \ForEach{$(s,o) \in L$}
    %      {

    \nl $o \gets \{ \l \in \Loc_1 \mid o' \cup (s' \cap \alpha_1) \models \delta_1(\l,\sigma) \}$ \label{alg:pre-alt-o}\;

    \nl \If{$o' \not\subseteq \alpha_1 \lor o' = \emptyset$ \label{alg:pre-alt-simulation-test}}
	{
	   \nl $L_{\sf Pre} \gets \{ \tuple{o,\emptyset} \}$ \label{alg:pre-alt-o-added}\; 		
	}

    \nl \If{$o \neq \emptyset$ \label{alg:pre-alt-line-emtpiness-test}}
        {
	    \nl $s \gets \{ \l \in \Loc_1 \mid s' \models \delta_1(\l,\sigma) \}$ \label{alg:pre-alt-s} \;
	    \nl $L_{\sf Pre} \gets L_{\Pre} \cup \{ \tuple{s,o} \}$ \label{alg:pre-alt-s-added}\;
	}
     %     }

    %}
    \nl \KwRet{$L_{{\sf Pre}}$}\; }\medskip

\caption{Algorithm for $\Pre^{{\sf alt}}_{\sigma}(\cdot)$. \label{alg:pre-alt}}

\end{algorithm}

\begin{thm}\label{theo:correctness-alg-pre-alt}
  Given an ABW $\A_1 = \tuple{\Loc_1, \iota_1, \Sigma, \delta_1, \alpha_1}$, 
  $\sigma \in \Sigma$ and $\tuple{s',o'} \in 2^{\Loc_1} \times 2^{\Loc_1}$ such that $o' \subseteq s'$, 
  the set $L_{\Pre} = \Pre^{{\sf alt}}_{\sigma}(\tuple{s,o})$
  computed by Algorithm~1 is an $\preceq_{{\sf alt}}$-antichain such that
  $\closure{L_{\Pre}}= \Pre^{\A_2}_{\sigma}(\closure{\{\tuple{s',o'}\}})$ where 
  $\A_2 = \MH(\A_1)$.
\end{thm}

\proof 
Let $\A_2 = \MH(\A_1) = \tuple{\Loc_2, \iota_2, \Sigma, \delta_2, \alpha_2}$.
The following entails that $\closure{L_{\Pre}}=
\Pre^{\A_2}_{\sigma}(\closure{\{\tuple{s',o'}\}})$:
\begin{enumerate}[(a)]
\item $L_{\Pre} \subseteq
  \Pre^{\A_2}_{\sigma}(\closure{\{\tuple{s',o'}\}})$, and 
\item[(b)] for all $\tuple{s_1,o_1} \in
  \Pre^{\A_2}_{\sigma}(\closure{\{\tuple{s',o'}\}})$, 
\end{enumerate}
there exists $\tuple{s,o} \in L_{\Pre}$ such that $\tuple{s_1,o_1} \preceq_{{\sf alt}} \tuple{s,o}$.

To prove (a), we first show that $\tuple{s,o} \xrightarrow{\sigma}_{\delta_2} \tuple{s',o'}$
where $\tuple{s,o}$ is added to $L_{\Pre}$ at line~\ref{alg:pre-alt-s-added} of 
Algorithm~1.
By the test of line~\ref{alg:pre-alt-line-emtpiness-test}, we have $o \neq \emptyset$. 
According to Definition~\ref{def:MH-construction} of $\MH(\cdot)$, we check that 
there exists a set $o'' \subseteq s'$ such that $o' = o'' \setminus \alpha_1$
(we take $o'' = o' \cup (s' \cap \alpha_1)$), and the following conditions hold:
\begin{enumerate}[$(i)$]
\item $s' \models \bigwedge_{\l \in s} \delta_1(\l,\sigma)$ since 
we have $s' \models \delta_1(\l,\sigma)$ for all $\l \in s$ 
by line~\ref{alg:pre-alt-s} of Alg.~1.%\\[-9pt]
\item $o'' \models \bigwedge_{\l \in o} \delta_1(\l,\sigma)$ since
we have $o'' \models \delta_1(\l,\sigma)$ for all $\l \in o$
by line~\ref{alg:pre-alt-o} of Alg.~1.%\\[-12pt]
\end{enumerate}

Second, we show that $\tuple{o,\emptyset} \xrightarrow{\sigma}_{\delta_2} \tuple{s'',o''}$ for 
some $\tuple{s'',o''} \preceq_{{\sf alt}} \tuple{s',o'}$ where $\tuple{o,\emptyset}$
is added to $L_{\Pre}$ at line~\ref{alg:pre-alt-o-added} of Algorithm~1.
We take $s'' = o' \cup (s' \cap \alpha_1)$ and $o'' = s'' \setminus \alpha_1$. 
Since $o' \subseteq s'$, we  have $(a)$ $s'' \subseteq s'$,
and we have $(b)$ $o'' = o' \setminus \alpha_1 \subseteq o'$. 
Let us establish that $(c)$ $o' = \emptyset$ iff $o'' = \emptyset$.
If $o' = \emptyset$ then $o'' = \emptyset$ since $o'' \subseteq o'$.
If $o' \neq \emptyset$ then by the test of line~\ref{alg:pre-alt-simulation-test},
we have $o' \not\subseteq \alpha_1$ and thus $o'' = o' \setminus \alpha_1 \neq \emptyset$.
Hence we have $\tuple{s'',o''} \preceq_{{\sf alt}} \tuple{s',o'}$,
and by line~\ref{alg:pre-alt-o} of the algorithm, we have 
$s'' \models \delta_1(\l,\sigma)$ for all $\l \in o$, and thus 
$s'' \models \bigwedge_{\l \in o} \delta_1(\l,\sigma)$. Therefore
$\tuple{o,\emptyset} \xrightarrow{\sigma}_{\delta_2} \tuple{s'',o''}$.

To prove (b), assume that there exist $\tuple{s_1,o_1}$ and $\tuple{s_1',o_1'}$
such that $\tuple{s_1,o_1} \xrightarrow{\sigma}_{\delta_2} \tuple{s_1',o_1'}$ and 
$\tuple{s_1',o_1'} \preceq_{{\sf alt}} \tuple{s',o'}$.
We have to show that there exists $\tuple{s,o} \in L_{\Pre}$ such that 
$\tuple{s_1,o_1} \preceq_{{\sf alt}} \tuple{s,o}$.

First, assume that $o_1 \neq \emptyset$.
Since $\tuple{s_1,o_1} \xrightarrow{\sigma}_{\delta_2} \tuple{s_1',o_1'}$,
we have:
\begin{enumerate}[$(i)$]  
\item for all $\l \in s_1$, $s_1' \models \delta_1(\l,\sigma)$ and since 
$s_1' \subseteq s'$ also $s' \models \delta_1(\l,\sigma)$. Let $s$
be the set defined at line~\ref{alg:pre-alt-s} of Algorithm~1.
For all $\l \in \Loc$, if $s' \models \delta_1(\l,\sigma)$ then $\l \in s$.
Hence, $s_1 \subseteq s$.

\item for all $\l \in o_1$, $o_1'' \models \delta_1(\l,\sigma)$
for some $o_1'' \subseteq s'_1$ such that $o_1' = o_1'' \setminus \alpha_1$. 
Hence necessarily $o_1'' \subseteq o_1' \cup (s_1' \cap \alpha_1) \subseteq o' \cup (s' \cap \alpha_1)$
and thus for all $\l \in o_1$, $o' \cup (s' \cap \alpha_1) \models \delta_1(\l,\sigma)$.
Let $o$ be the set defined at line~\ref{alg:pre-alt-o} of Algorithm~1.
For all $\l \in \Loc$, if $o' \cup (s' \cap \alpha_1) \models \delta_1(\l,\sigma)$ 
then $\l \in o$. Hence, $o_1 \subseteq o$ and $o \neq \emptyset$.
\end{enumerate}
\noindent
Hence, $\tuple{s,o}$ which is added to $L_{\Pre}$ by Alg.~1 at 
line~\ref{alg:pre-alt-s-added} satisfies $\tuple{s_1,o_1} \preceq_{{\sf alt}} \tuple{s,o}$.

Second, assume that $o_1 = \emptyset$. 
Since $\tuple{s_1,o_1} \xrightarrow{\sigma}_{\delta'} \tuple{s_1',o_1'}$ and $o_1 = \emptyset$,
we know that for all $\l \in s_1$, $s_1' \models \delta_1(\l,\sigma)$ and 
$o_1' = s_1' \setminus \alpha_1$. Let $s'' = o' \cup (s' \cap \alpha_1)$ so we have 
$(a)$ $s_1' \cap \alpha_1 \subseteq s' \cap \alpha_1 \subseteq s''$ and 
$(b)$ $s_1' \setminus \alpha_1 = o_1' \subseteq o'  \subseteq s''$.
Hence, $s_1' \subseteq s''$ and thus for all $\l \in s_1$, $s'' \models \delta_1(\l,\sigma)$.
Let $o$ be the set defined at line~\ref{alg:pre-alt-o} of Algorithm~1.
For all $\l \in \Loc$, if $s'' \models \delta_1(\l,\sigma)$ 
then $\l \in o$. Hence, $s_1 \subseteq o$ and 
$\tuple{s_1,\emptyset} \preceq_{{\sf alt}} \tuple{o, \emptyset}$ where $\tuple{o, \emptyset}$
is added to $L_{\Pre}$ by Algorithm~1 at line~\ref{alg:pre-alt-o-added}.
Notice that the test at line~\ref{alg:pre-alt-simulation-test} is satisfied because
$o_1' = s_1' \setminus \alpha_1$ implies that $o_1' \not\subseteq \alpha_1 \lor 
o_1' = \emptyset$ and since $\tuple{s_1',o_1'} \preceq_{{\sf alt}} \tuple{s',o'}$,
we have $o' \not\subseteq \alpha_1 \lor o' = \emptyset$.
\qed

% Section 5

\section{Universality of NBW}\label{sec:universality}

We present a new algorithm to check universality of NBW, based
the existence of a simple simulation relation for the complement
automaton of NBW given by Definition~\ref{def:direct-complementation}.

\begin{defi}\label{def:simulation-for-KVMH}
Given an NBW $\A = \tuple{\Loc, \iota, \Sigma, \delta, \alpha}$,
let $\KVMH(\A) = \tuple{Q \times Q, q_{\iota}, \Sigma, \delta', \alpha'}$.
Define the pre-order $\preceq_{{\sf univ}}\, \subseteq (Q \times Q) \times (Q \times Q)$ as follows: 
for all $s,s',o,o' \in Q$, let 
$\tuple{s,o} \preceq_{{\sf univ}} \tuple{s',o'}$ 
iff the following conditions hold:
  \begin{enumerate}[$\bullet$]
    \item for all $(\l,n) \in s$, there exists $n' \leq n$ such that $(\l,n') \in s'$;
    \item for all $(\l,n) \in o$, there exists $n' \leq n$ such that $(\l,n') \in o'$;
    \item $o=\emptyset$ iff $o'=\emptyset$.
  \end{enumerate}
\end{defi}

This relation formalizes the intuition that it is easier to accept a word 
in $\KVMH(\A)$ from a given location with a high rank than with a small rank. 
This is because the rank is always decreasing along every path of the runs 
of $\KV(\A)$, and so a small rank is always simulated by a greater rank.
Hence, essentially the minimal rank of each location of $s$ and $o$ is relevant
to define the pre-order $\preceq_{{\sf univ}}$. The third condition requires
that only accepting states simulate accepting states.

% The relation $\preceq_{{\sf univ}}$ is a simulation
% for the NBW $\KVMH(\A,k)$ (with state space $Q_k \times Q_k$) defined 
% in Section~\ref{sec:direct-complementation}.

\begin{lem}\label{lem:univ-simulation-relation}
For all NBW $\A$, 
the pre-order $\preceq_{{\sf univ}}$ is a simulation for the NBW $\KVMH(\A)$.
\end{lem} 

% The proof of Lemma~\ref{lem:univ-simulation-relation} is given in the appendix.
\proof %[of Lemma~\ref{lem:univ-simulation-relation}]
Let $\A = \tuple{\Loc, \iota, \Sigma, \delta, \alpha}$ and 
$\KVMH(\A) = \tuple{Q \times Q, q_{\iota}, \Sigma, \delta', \alpha'}$.
First, we show that for all 
$\tuple{s_1,o_1}, \tuple{s_2,o_2}, \tuple{s_3,o_3} \in Q \times Q$, for all $\sigma \in \Sigma$,
if $\tuple{s_1,o_1} \xrightarrow{\sigma}_{\delta'} \tuple{s_2,o_2}$ 
and $\tuple{s_3,o_3} \preceq \tuple{s_1,o_1}$ 
then $\tuple{s_3,o_3} \xrightarrow{\sigma}_{\delta'} \tuple{s_2,o_2}$. Notice that we have trivially 
$\tuple{s_2,o_2} \preceq_{{\sf univ}} \tuple{s_2,o_2}$.
We give the proof for $o_1 \neq \emptyset$. 
The case $o_1 = \emptyset$ is proven similarly.
According to Definition~\ref{def:direct-complementation}, 
since $\tuple{s_1,o_1} \xrightarrow{\sigma}_{\delta'} \tuple{s_2,o_2}$ we have
% $(i)$ $\forall (\l,n_1) \in s_1 \cdot \forall \l' \in \delta(\l,\sigma)\cdot \exists n_2 \leq n_1: (\l',n_2) \in s_2$
% and $(ii)$ $\forall (\l,n_1) \in o_1 \cdot \forall \l' \in \delta(\l,\sigma)\cdot \exists n_2 \leq n_1: (\l',n_2) \in o_2$.
$$ \begin{array}{rl}
$(i)$ & \forall (\l,n_1) \in s_1 \cdot \forall \l' \in \delta(\l,\sigma)\cdot \exists n_2 \leq n_1: (\l',n_2) \in s_2 \text{ and}\\
$(ii)$& \forall (\l,n_1) \in o_1 \cdot \forall \l' \in \delta(\l,\sigma)\cdot \exists n_2 \leq n_1: (\l',n_2) \in o_2
\end{array} 
$$
Since $\tuple{s_3,o_3} \preceq \tuple{s_1,o_1}$, we have $o_3 \neq \emptyset$ and
% $(i')$ $\forall (\l,n_3) \in s_3 \cdot \exists (\l,n_1) \in s_1: n_1 \leq n_3$
% and $(ii')$ $\forall (\l,n_3) \in o_3 \cdot \exists (\l,n_1) \in o_1: n_1 \leq n_3$.
$$ \begin{array}{rl}
(i') & \forall (\l,n_3) \in s_3 \cdot \exists n_1 \leq n_3: (\l,n_1) \in s_1 \text{ and}\\
(ii')& \forall (\l,n_3) \in o_3 \cdot \exists n_1 \leq n_3: (\l,n_1) \in o_1
\end{array} 
$$
Combining $(i)$ and $(i')$ yields 
$\forall (\l,n_3) \in s_3 \cdot \forall \l' \in \delta(\l,\sigma)\cdot \exists n_2  \leq n_3: (\l',n_2) \in s_2:$, 
and combining $(ii)$ and $(ii')$ yields
$\forall (\l,n_3) \in o_3 \cdot \forall \l' \in \delta(\l,\sigma)\cdot \exists n_2 \leq n_3: (\l',n_2) \in o_2$. 
Since $o_3 \neq \emptyset$, this implies that $\tuple{s_3,o_3} \xrightarrow{\sigma}_{\delta'} \tuple{s_2,o_2}$.

Second, for all $\tuple{s,o} \in \alpha'$ we have $o = \emptyset$, and
thus for all $\tuple{s',o'} \in Q \times Q$, if
$\tuple{s',o'} \preceq \tuple{s,o}$ then $o' = \emptyset$ so that
$\tuple{s',o'} \in \alpha'$.

Hence $\preceq_{{\sf univ}}$ is a simulation for $\KVMH(\A)$.
\qed

According to Lemmas~\ref{lem:closure-properties-a} and~\ref{lem:closure-properties-b}, 
all intermediate sets that are computed by the fixed point $\F_{\A^c}$ to check emptiness of
$\A^c = \KVMH(\A)$ (and thus universality of~$\A$) are $\preceq_{{\sf univ}}$-closed. 
Since~$\preceq_{{\sf univ}}$ is not a partial order, 
the set $\Max(L)$ for a $\preceq_{{\sf univ}}$-closed set~$L$
may contain several $\preceq_{{\sf univ}}$-equivalent elements ($x$ and $y$ are 
$\preceq_{{\sf univ}}$-equivalent if $x \preceq_{{\sf univ}} y$ and $y \preceq_{{\sf univ}} x$).
For example, the set $\{\tuple{\{(\l,3),(\l',4)\},\emptyset}\}$ is 
$\preceq_{{\sf univ}}$-equivalent to the set 
$\{\tuple{\{(\l,3),(\l,4),(\l',4)\},\emptyset}\}$.
In fact $\Max(L)$ is a union of $\preceq_{{\sf univ}}$-equivalent classes.
Hence, the size of $\Max(L)$ can be reduced by keeping only one canonical 
element for each $\preceq_{{\sf univ}}$-equivalent class.
Given a set $s \in Q$, define its \emph{characteristic function} $f_s:\Loc \to \nat
\cup \{ \infty \}$ such that $f_s(\l) = \inf\{n \mid (\l,n) \in s\}$
with the usual convention that $\inf\; \emptyset = \infty$.
Note that if $f_s(\l) \neq \infty$, then $f_s(\l)$ is even for all $\l \in \alpha$.

Let $f,g,f',g'$ be characteristic functions. Let $\max(f,f')$ be the function~$f''$ 
such that $f''(\l) = \max\{f(\l), f'(\l)\}$ for all $\l \in \Loc$.
We denote by $f_{\emptyset}$ the function such that $f_{\emptyset}(\l) = \infty$ for all $\l \in \Loc$.  
We write $f \leq f'$ if for all $\l \in \Loc$, $f(\l) \leq f'(\l)$ and we write 
$\tuple{f,g} \leq \tuple{f',g'}$ if $f \leq f'$, $g \leq g'$ and $g = f_{\emptyset}$ iff $g = f_{\emptyset}$.
Notice that~$\leq$ is partial order over characteristic functions,
and that if $s \subseteq s'$, then $f_{s'} \leq f_s$ for all $s,s' \in Q$.
The following lemma is a corollary of Definition~\ref{def:simulation-for-KVMH}.

\begin{lem}\label{lem:characteristic-function}
  For all sets $s,s',o,o' \in Q$, $\tuple{f_{s'},f_{o'}} \leq  \tuple{f_{s},f_{o}}$
  if and only if $\tuple{s,o} \preceq_{{\sf univ}} \tuple{s',o'}$.
\end{lem}

Define $\sem{f} = \{s \in Q \mid \exists s' \in Q: s \subseteq s' \land f_{s'} = f\}$ and 
$\sem{\tuple{f,g}} = \{\tuple{s,o} \mid \tuple{f,g} \leq  \tuple{f_{s},f_{o}} \}$.
We extend the operator $\sem{\cdot}$ to sets of pairs of characteristic
functions as expected. 
Notice that $f \leq f'$ iff $\sem{f'} \subseteq \sem{f}$, that $\sem{\max(f,f')} = \sem{f} \cap \sem{f'}$, and 
a corollary of Lemma~\ref{lem:characteristic-function} is that 
the $\leq$-minimal elements of a set $L$ of pairs of characteristic functions
represents exactly the $\preceq_{{\sf univ}}$-maximal pairs $\tuple{s,o}$ of $\sem{L}$.

% A corollary of Lemma~\ref{lem:characteristic-function} is that if 
% $\tuple{f_{s'},f_{o'}} =  \tuple{f_{s},f_{o}}$, then $\tuple{s,o}$ and $\tuple{s',o'}$
% are $\preceq_{{\sf univ}}$-equivalent.
% Therefore, the set $\sem{\tuple{f,g}}$ is an equivalence class of the equivalence 
% relation induced by $\preceq_{{\sf univ}}$. 
% Since the $\preceq_{{\sf univ}}$-closed sets (as
% well as their $\preceq_{{\sf univ}}$-maximal elements) are unions of
% $\preceq_{{\sf univ}}$-equivalence classes, they can be (more compactly) 
% viewed as unions of pairs of characteristic functions.

Now, we show how to compute efficiently $\cup$, $\cap$ and $\Pre$ for
$\preceq_{{\sf univ}}$-closed sets that are represented by
characteristic functions. 
Let $L_1, L_2$ be two sets of pairs of
characteristic functions, let $L_{\cup}$ be the set
of $\leq$-minimal elements of $L_1 \cup L_2$, and let $L_{\cap}$ be the $\leq$-minimal elements
of the union of:
\begin{enumerate}[$\bullet$]
\item[] $\{\tuple{\max(f_s,f_{s'}), \max(f_o,f_{o'})} \mid 
\tuple{f_s,f_o} \in L_1 \land \tuple{f_{s'},f_{o'}} \in L_2 \land \max(f_o,f_{o'}) \not= f_{\emptyset} \}$ and
\item[] $\{\tuple{\max(f_s,f_{s'}), f_{\emptyset}} \mid 
\tuple{f_s,f_{\emptyset}} \in L_1 \land \tuple{f_{s'},f_{\emptyset}} \in L_2 \}$.
\end{enumerate}
By Equation~\eqref{eq:intersection} and by the previous remarks, we have:
\begin{enumerate}[$\bullet$]
\item[] $\sem{L_{\cup}} = \sem{L_1} \cup \sem{L_2}$ and $\sem{L_{\cap}} = \sem{L_1} \cap \sem{L_2}$.
\end{enumerate}\medskip

% In the sequel, we often write $\Pre^{{\sf univ}}_{\sigma}$ instead of $\Pre^{\KVMH(\A,k)}_{\sigma}$
% when $\A$ and $k$ are clear from the context.
\noindent To compute $\Pre_{\sigma}(\cdot)$ of a single pair of
characteristic functions, we propose Algorithm~2 whose correctness is
established by Theorem~\ref{theo:correctness-alg-pre}. Computing the
predecessors of a set of characteristic functions is then
straightforward using the algorithm for union of sets of pairs of
characteristic functions since
$$\Pre^{\KVMH(\A)}(L) = \bigcup_{\sigma \in \Sigma} \bigcup_{\l \in L} \Pre^{\KVMH(\A)}_{\sigma}(\l).$$
In Algorithm~2, we represent $\infty$ by any number
strictly greater than $k = 2(\abs{\Loc} - \abs{\alpha})$, and we adapt the definition of $\leq$ as
follows: $f \leq f'$ iff for all $\l \in \Loc$, either $f(\l) \leq
f'(\l)$ or $f'(\l) > k$. In the algorithm, we use the notations
$\ceilOdd{n}$ for the least odd number $n'$ such that $n' \geq n$, and
$\ceilEven{n}$ for the least even number $n'$ such that $n' \geq n$.

The structure of Algorithm~2
is similar to Algorithm~1, but the computations are expressed 
in terms of characteristic functions, thus in terms of ranks. For example,
\mbox{lines~\ref{alg:pre-line-o-alpha}-\ref{alg:pre-line-o-not-alpha}} compute
the equivalent of line~\ref{alg:pre-alt-o} in Algorithm~1,
where $\alpha_1$ corresponds here to the set of odd-ranked locations,
and thus contains no $\alpha$-nodes. Details are given in the proof of 
Theorem~\ref{theo:correctness-alg-pre}.

\begin{thm}\label{theo:correctness-alg-pre}
  Let $\A = \tuple{\Loc, \iota, \Sigma, \delta, \alpha}$ be a NBW, 
  $\sigma \in \Sigma$, and $\tuple{f_{s'},f_{o'}}$ be a pair of characteristic functions 
  such that $f_{s'} \leq f_{o'}$.
  The set $L_{\Pre} = \Pre^{{\sf univ}}_{\sigma}(\tuple{f_{s'},f_{o'}})$ computed by
  Algorithm~2 is such that 
  $\sem{L_{\Pre}} = \Pre^{\KVMH(\A)}_{\sigma}(\sem{\tuple{f_{s'},f_{o'}}})$
  and for all $\tuple{f_{s},f_{o}} \in L_{\Pre}$, we have $f_{s} \leq f_{o}$
  and $f_{s}(\l)$ and $f_{o}(\l)$ are even for all $\l \in \alpha$.
\end{thm} 

%The proof of Theorem~\ref{theo:correctness-alg-pre} is given in the appendix.
\proof 
Let $\A^c = \KVMH(\A) = \tuple{Q \times Q, q_{\iota}, \Sigma, \delta', \alpha'}$,
and let $\tuple{s',o'}$ be a pair of sets whose characteristic functions
are $\tuple{f_{s'},f_{o'}}$ and $o'\subseteq s'$ (such a pair exists because
$f_{s'} \leq f_{o'}$).
We show that (a) $\sem{L_{\Pre}} \subseteq \Pre^{\A^c}_{\sigma}(\sem{\tuple{f_{s'},f_{o'}}})$ and
(b) $\Pre^{\A^c}_{\sigma}(\sem{\tuple{f_{s'},f_{o'}}}) \subseteq \sem{L_{\Pre}}$.

To prove (a), first consider a pair $\tuple{f_s,f_o}$ added to $L_{\Pre}$ at 
line~\ref{alg:pre-line-Z-one} of Algorithm~2 and let 
$\tuple{s,o} \in \sem{\tuple{f_s,f_o}}$. We show that 
$\tuple{s,o} \xrightarrow{\sigma}_{\delta'} \tuple{s',o'}$ and $f_s \leq f_o$.

By the test of line~\ref{alg:pre-line-emtpiness-test}, we have $f_o \neq f_{\emptyset}$
and therefore $o \neq \emptyset$. 
According to Definition~\ref{def:direct-complementation} of $\KVMH(\A)$, 
we have to check that there exists a set $o'' \subseteq s'$ such that 
$o' = o'' \setminus \odd$ (we take $o'' = o' \cup (s' \cap \odd)$), 
and the following conditions hold:
\begin{enumerate}[$(i)$]
\item $\forall (\l,n) \in s \cdot \forall \l' \in \delta(\l,\sigma)\cdot \exists n' \leq n: (\l',n') \in s'$.

		Observe that for all $\l \in \Loc$, for all $\l' \in \delta(\l,\sigma)$, 
		we have $f_{s'}(\l') \leq f_s(\l)$ (lines~\ref{alg:pre-line-f-s},\ref{alg:pre-line-s-ceileven} 
		of Algorithm~2). 
		Since $f_s(\l) \leq n$ (by definition of characteristic functions), 
		we take $n' = f_{s'}(\l')$ so that we have $n' \leq f_s(\l) \leq n$ and $(\l',n') \in s'$. 
		
\item $\forall (\l,n) \in o \cdot \forall \l' \in \delta(\l,\sigma)\cdot \exists n' \leq n: (\l',n') \in o''$.

		Since $o'' = o' \cup (s' \cap \odd)$, we have $f_{o''}(\l') = f_{o'}(\l')$ for $\l' \in \alpha$
		and $f_{o''}(\l') = \min\{ f_{o'}(\l'), \ceilOdd{f_{s'}(\l')} \}$ for $\l' \not\in \alpha$.
		Now, for all $\l \in \Loc$, for all $\l' \in \delta(\l,\sigma)$, 
		we have either $\l' \in \alpha$ and then $f_o(\l) \geq n'$ for $n' = f_{o'}(\l')$, 
		or $\l' \not\in \alpha$ and then $f_o(\l) \geq n'$ for $n' = \min\{ f_{o'}(\l'), \ceilOdd{f_{s'}(\l')} \}$
		(lines~\ref{alg:pre-line-o-alpha}-\ref{alg:pre-line-o-ceileven} 
		of Algorithm~2).
		In both cases, for $(\l,n) \in o$ we have $f_{o''}(\l') \leq n' \leq f_o(\l) \leq n$ and $(\l',n') \in o''$.
\end{enumerate}
Moreover, we prove that:
\begin{enumerate}[$(i)$]
\item[$(iii)$] $f_s \leq f_o$.

Since $f_{s'} \leq f_{o'}$, we have for all $\l' \in \Loc$ either 
$f_{o'}(\l') > k$ or $f_{o'}(\l') \geq f_{s'}(\l')$.
By lines~\ref{alg:pre-line-o-alpha}-\ref{alg:pre-line-o-ceileven} of Algorithm~2,
we have for all $\l \in \Loc$, for all $\l' \in \delta(\l,\sigma)$ 
either $f_o(\l) \geq f_{o'}(\l')$ or $f_o(\l) \geq \ceilOdd{f_{s'}(\l')}$,
and thus either $f_o(\l) > k$ or $f_o(\l) \geq f_{s'}(\l')$. 
Hence, we have for all $\l \in \Loc$ either $f_o(\l) > k$ or 
$f_o(\l) \geq \max \{f_{s'}(\l') \mid \l' \in \delta(\l,\sigma) \}$.
Therefore, by lines~\ref{alg:pre-line-f-s}-\ref{alg:pre-line-s-ceileven} of 
Algorithm~2, if $\l \not\in \alpha$, then $f_o(\l) > k$ or $f_o(\l) \geq f_{s}(\l)$,
and if $\l \in \alpha$, then $f_o(\l)$ is even (line~\ref{alg:pre-line-o-ceileven})
and thus either $f_o(\l) > k$ or $f_o(\l) \geq \ceilEven{\max \{f_{s'}(\l') \mid \l' \in \delta(\l,\sigma) \}} = f_{s}(\l)$.
In all cases, $f_s \leq f_o$.

\item[(iv)] $\forall \l \in \alpha: f_s(\l)$ and $f_o(\l)$ are even. 

This is enforced by line~\ref{alg:pre-line-s-ceileven} and line~\ref{alg:pre-line-o-ceileven} of the algorithm.
\end{enumerate}

\noindent Second, consider a pair $\tuple{f_o,f_{\emptyset}}$ added to
$L_{\Pre}$ at line~\ref{alg:pre-line-Z-two}, and let
$\tuple{s,\emptyset} \in \sem{\tuple{f_o,f_{\emptyset}}}$.  Notice
that $f_o \leq f_{\emptyset}$ and that $f_o(\l)$ is even for all $\l
\in \alpha$ by $(iv)$.  We show that there exists $\tuple{s'',o''}
\preceq_{{\sf univ}} \tuple{s',o'}$ such that $\tuple{s,\emptyset}
\xrightarrow{\sigma}_{\delta'} \tuple{s'',o''}$.  We take $s'' = o'
\cup (s' \cap \odd)$ and $o'' = s'' \setminus \odd$.  Since $o'
\subseteq s'$, we have $(1)$ $s'' \subseteq s'$, and we have $(2)$
$o'' = o' \setminus \odd \subseteq o'$.  Moreover, if $o' \neq
\emptyset$, then there exists let $(\l,n) \in o'$ for some $\l \in
\Loc$ and even number $n$, since the maximal rank $k = 2(\abs{\Loc} -
\abs{\alpha})$ is even. So $(\l,n) \in o''$ and thus $o'' \neq
\emptyset$.  Since $o'' \subseteq o'$, we have $(3)$ $o' \neq
\emptyset$ iff $o'' \neq \emptyset$.  Hence $\tuple{s'',o''}
\preceq_{{\sf univ}} \tuple{s',o'}$. The fact that
$\tuple{f_o,\emptyset} \xrightarrow{\sigma}_{\delta'} \tuple{s'',o''}$
is proven similarly to $(ii)$.

To prove (b), assume that there exist $\tuple{s_1,o_1}$ and $\tuple{s_1',o_1'}$
such that $\tuple{s_1,o_1} \xrightarrow{\sigma}_{\delta'} \tuple{s_1',o_1'}$ and 
$\tuple{s_1',o_1'} \in \sem{\tuple{f_{s'},f_{o'}}}$.
We have to show that $\tuple{s_1,o_1} \in \sem{L_{\Pre}}$, i.e., 
$\tuple{f_{s_1},f_{o_1}} \geq \tuple{f_s,f_o}$ for some $\tuple{f_s,f_o} \in L_{\Pre}$.

First, assume that $o_1 \neq \emptyset$. 
Notice that $f_{s_1'} \geq f_{s'}$ and $f_{o_1'} \geq f_{o'}$ 
since $\tuple{s_1',o_1'} \in \sem{\tuple{f_{s'},f_{o'}}}$,
From the fact that $\tuple{s_1,o_1} \xrightarrow{\sigma}_{\delta'} \tuple{s_1',o_1'}$,
we get:
\begin{enumerate}[$(i)$] 
\item for all $(\l,n_1) \in s_1$, for all $\l' \in \delta(\l,\sigma)$, $n_1 \geq f_{s_1}(\l) \geq f_{s_1'}(\l')$ 
and thus $n_1 \geq f_{s'}(\l')$. Hence, for all $\l \in \Loc$ we have
$f_{s_1}(\l) \geq \max\{f_{s'}(\l') \mid \l' \in \delta(\l,\sigma)\} = f_s(\l)$,
where $f_s$ is computed by line~\ref{alg:pre-line-f-s} of Algorithm~2) for $\l \not\in \alpha$. 
We also have $f_{s_1}(\l) \geq f_s(\l)$ (see line~\ref{alg:pre-line-s-ceileven} of Algorithm~2) 
for $\l \in \alpha$, as $f_{s_1}(\l)$ is even in that case.
Thus, $f_s \leq f_{s_1}$.
 
\item for all $(\l,n_2) \in o_1$, for all $\l' \in \delta(\l,\sigma)$, $n_2 \geq f_{o_1}(\l) \geq f_{o_1''}(\l')$ 
for some set $o_1''$ such that $o_1'' \subseteq s_1'$ and $o_1'' \setminus \odd = o_1'$. 
Therefore $o_1'' \subseteq o_1' \cup (s_1' \cap \odd)$
and thus $f_{o_1''} \geq f_{o_1' \cup (s_1' \cap \odd)} \geq f_{o' \cup (s' \cap \odd)}$
since $f_{s_1'} \geq f_{s'}$ and $f_{o_1'} \geq f_{o'}$.
Hence, for all $\l \in \Loc$ 
either $f_{o_1}(\l) > k$ or $f_{o_1}(\l) \geq f_o(\l)$ (where $f_o$ is computed 
at lines~\ref{alg:pre-line-f-o-begin}-\ref{alg:pre-line-f-o-end} of Algorithm~2).
Thus, $f_o \leq f_{o_1}$.

\item By our assumption that $o_1 \neq \emptyset$, we have $f_{o_1} \neq f_{\emptyset}$, and 
so $f_{o} \neq f_{\emptyset}$ by $(ii)$.
\end{enumerate}\medskip

\noindent Hence, the pair $\tuple{f_s,f_o}$ added to $L_{\Pre}$ by
Algorithm~2 at line~\ref{alg:pre-line-Z-one} satisfies
$\tuple{f_{s_1},f_{o_1}} \geq \tuple{f_s,f_o}$ and thus
$\tuple{s_1,o_1} \in \sem{L_{\Pre}}$.

% and if $\tuple{f_s,f_o}$ is later removed from $L_{\Pre}$, it must have been
% replaced by a $\leq$-smaller one, and thus also $\preceq_{{\sf univ}}$-greater 
% than $\tuple{s_1,o_1}$.

Second, assume that $o_1 = \emptyset$. Let $s'' = o' \cup (s' \cap \odd)$. 
Since $\tuple{s_1,o_1} \xrightarrow{\sigma}_{\delta'} \tuple{s_1',o_1'}$ and $o_1 = \emptyset$,
we have $o_1' = s_1' \setminus \odd$. 
Next, we use several times the fact that $u \subseteq v$ implies $f_v \leq f_u$. 
Since $f_{s_1'} \geq f_{s'}$ and $f_{o_1'} \geq f_{o'}$, we have 
$(1)$ $f_{s_1' \cap \odd} \geq f_{s' \cap \odd} \geq f_{s''}$ 
and $(2)$ $f_{s_1' \setminus \odd} = f_{o_1'} \geq f_{o'} \geq f_{s''}$. 
By $(1)$ and $(2)$, we get easily $f_{s_1'} \geq f_{s''}$.
Now, by the fact that 
$\tuple{s_1,o_1} \xrightarrow{\sigma}_{\delta'} \tuple{s_1',o_1'}$, we know that
for all $(\l,n_1) \in s_1$, for all $\l' \in \delta(\l,\sigma)$, $n_1 \geq f_{s_1'}(\l')$ 
and thus $n_1 \geq f_{s''}(\l')$. 
Notice that $f_o(\l) = \max\{f_{s''}(\l') \mid \l' \in\delta(\l,\sigma)\}$,
where $f_o$ is computed at lines~\ref{alg:pre-line-f-o-begin}-\ref{alg:pre-line-f-o-end} of Algorithm~2. 
Thus, $n_1 \geq f_{o}(\l)$ for all $\l \in \Loc$ 
and therefore $f_{s_1} \leq f_{o}$ so that $\tuple{s_1,o_1} \in \sem{\tuple{f_o,f_{\emptyset}}}$
where $\tuple{f_o,f_{\emptyset}}$ is added to $L_{\Pre}$ by Algorithm~2 at line~\ref{alg:pre-line-Z-one}. 
% If it is later removed from $L_{\Pre}$, it must have been replaced by a $\leq$-smaller one, 
% and thus also $\preceq_{{\sf univ}}$-greater than $\tuple{s_1,\emptyset}$.
\qed

Algorithm~2 computes the predecessors of a pair $\tuple{f_{s'},f_{o'}}$
in time $O(\abs{\Loc}^2)$, which is polynomial in the size of the input, even though
the number of pairs $\tuple{s',o'}$ that are represented by the pair $\tuple{f_{s'},f_{o'}}$
and by the computed set $L_{\Pre}$ can be of exponential size.
For example, the set $\alpha'= Q \times \{\emptyset\}$ 
with an exponential number of elements is represented by the unique
pair $\tuple{f_s,f_{\emptyset}}$ where $f_s(\l) = 0$ for all $\l \in \Loc$.
Hence the compact representation that we propose does not come
with an execution time blow-up, which makes the new approach much more efficient 
in practice.

\begin{algorithm}[!tbp]
  % {\scriptsize \SetKwComment{\tcc}{//}{}
  \AlgData{A NBW $\A = \tuple{\Loc, \iota, \Sigma, \delta, \alpha}$, $\sigma \in \Sigma$,
   and a pair $\tuple{f_{s'},f_{o'}}$ of characteristic functions.} 
  \AlgResult{The set $\Pre^{{\sf univ}}_{\sigma}(\tuple{f_{s'},f_{o'}})$.}
  \flushleft
  \Begin { 
    % \nl $k \gets 2(\abs{\Loc} - \abs{\alpha})$ \; 
    % \nl $L_{\Pre} \gets \emptyset$ \; 
    %\nl \ForEach{$\sigma \in \Sigma$ \label{alg:pre-line-for-sigma}}
    %{
    	\nl \ForEach{$\l \in \Loc$ \label{alg:pre-line-f-o-begin}}
	{
		\nl $f_o(\l) \gets 0$ \;
	    	\nl \ForEach{$\l' \in \delta(\l,\sigma)$}
		{
			\nl \lIf{$\l' \in \alpha$}{$f_o(\l) \gets \max \{f_o(\l), f_{o'}(\l')\}$} \label{alg:pre-line-o-alpha} \;
			\nl \lElse{$f_o(\l) \gets \max \{f_o(\l), \min\{f_{o'}(\l'), \ceilOdd{f_{s'}(\l')}\}\}$} \label{alg:pre-line-o-not-alpha}\;
		}
		\nl \lIf{$\l \in \alpha$}{$f_o(\l) \gets \ceilEven{f_o(\l)}$ \label{alg:pre-line-o-ceileven} \label{alg:pre-line-f-o-end}} \;
	}
	\nl $L_{\Pre} \gets \{\tuple{f_o,f_{\emptyset}}\}$ \label{alg:pre-line-Z-two}\;
	\nl $k \gets 2(\abs{\Loc} - \abs{\alpha})$ \;
	\nl \If{$\exists \l: f_o(\l) \leq k$ ({\it i.e.} $o \neq \emptyset$) \label{alg:pre-line-emtpiness-test}}
	{
		\nl \ForEach{$\l \in \Loc$}
		{
			\nl $f_s(\l) \gets \max \{f_{s'}(\l') \mid \l' \in \delta(\l,\sigma) \}$ \label{alg:pre-line-f-s} \;
			\nl \lIf{$\l \in \alpha$}{$f_s(\l) \gets \ceilEven{f_s(\l)}$ \label{alg:pre-line-s-ceileven}} \;
			
		}
		\nl $L_{\Pre} \gets L_{\Pre} \cup \{\tuple{f_s,f_o}\}$ \label{alg:pre-line-Z-one}\;
	}
	
    %}
 
  \nl \KwRet{$L_{\Pre}$}\; }\medskip

\caption{Algorithm for $\Pre^{{\sf univ}}_{\sigma}(\cdot)$. \label{alg:pre}}

\end{algorithm}

% Section 4

% Section 5

% Section 6
\section{Implementation and Practical Evaluation}\label{sec:implementation}

\paragraph{{\bf The randomized model}}
To evaluate our new algorithm for universality of NBW and compare with
the existing implementations of the Kupferman-Vardi and Miyano-Hayashi
constructions, we use a random model to generate NBW.  This model was
first proposed by Tabakov and Vardi to compare the efficiency of
some algorithms for automata in the context of finite words 
automata~\cite{TabakovV05} and more recently in the context of infinite
words automata~\cite{TabakovV07}.
In the model, the input alphabet is fixed to $\Sigma=\{0,1\}$, and for
each letter $\sigma \in \Sigma$, a number $k_\sigma$ of different
state pairs $(\l, \l') \in \Loc \times \Loc$ are chosen uniformly at
random before the corresponding transitions $(\l,\sigma,\l')$ are
added to the automaton.  The ratio $r_\sigma=\frac{k_\sigma}{|\Loc|}$
is called the \emph{transition density} for~$\sigma$.  This ratio
represents the average outdegree of each state for~$\sigma$.  In all
experiments, we choose $r_0=r_1$, and denote the transition density
by~$r$.  The model contains a second parameter: the \emph{density $f$
  of accepting states}.  There is only one initial state, and the
number $m$ of accepting states is linear in the total number of
states, as determined by $f=\frac{m}{|\Loc|}$.  The accepting states
themselves are chosen uniformly at random.  Observe that since the
transition relation is not always total, automata with $f=1$ are not
necessarily universal.

Tabakov and Vardi have studied the space of parameter values for this
model and argue that ``interesting'' automata are generated by the
model as the two parameters $r$ and $f$ vary. They also study the 
density of universal automata.

\paragraph{{\bf Performance comparison}}
We have implemented our algorithm to check the universality of
randomly generated NBW. The code is written in {\tt C} with an
explicit representation for characteristic functions, as arrays of
integers.  All the experiments are conducted on a biprocessor Linux
station (two $3.06$Ghz Intel Xeons with $4$GB of RAM).

\figurename~\ref{fig:median-time} shows as a function of $r$
(transition density) and $f$ (density of accepting states) the median
execution times for testing universality of $100$ random automata with
$\abs{\Loc}=30$.  It shows that the universality test was the most
difficult for $r=1.8$ and $f=0.1$ with a median time of $11$ seconds.
The times for $r \leq 1$ and $r \geq 2.8$ are not plotted because they
were always less than $250$ms.  A similar shape and maximal median
time is reported by Tabakov for automata
of size $6$, that is for automata that are five times smaller~\cite{TabakovV07}.  %%%%%% see page 19 of 29
Another previous work reports prohibitive execution times for
complementing NBW of size $6$, showing that explicitly constructing
the complement is not a reasonable approach~\cite{GurumurthyKSV03}.
The density of universal automata in the samples is shown in \figurename~\ref{fig:density}.
The density increases when states have more transitions, while 
it seems less sensitive to the density of accepting states. The difficult 
instances correspond to the values of the densities of transitions and 
accepting states for which the probability to be universal is close to a half.
Analogous results have been observed in~\cite{TabakovV07}.

To evaluate the scalability of our algorithm, 
we have run the following experiment. For a set of
parameter values, we have evaluated the maximal size of automata
(measured in term of number of locations) for which our algorithm 
could analyze $50$ over $100$ instances in less than 20 seconds. 
We have tried automata sizes from $10$ to $1500$, with a fine granularity
for small sizes (from $10$ to $100$ with an increment of $10$, from 
$100$ to $200$ with an increment of $20$, and from $200$ to $500$ 
with an increment of $30$) and a rougher granularity for
large sizes (from $500$ to $1000$ with an increment of $50$, and
from $1000$ to $1500$ with an increment of $100$).

\begin{figure}[!tbp]
  \begin{minipage}[b]{.45\linewidth}
    \scalebox{0.5}{\input{figures/TimeMedian-30-r0.20-3.00-0.20-f0.10-0.90-0.20-zoom.pstex}}
    \caption{\protect\parbox[t]{45mm}{Median time to check universality of $100$ automata of size $30$ for each sample point.}}
    \label{fig:median-time}
 \end{minipage} \hfill 
 \begin{minipage}[b]{.47\linewidth}
    \scalebox{0.5}{\input{figures/UAvg-30-30.pstex}}
    \caption{\protect\parbox[t]{45mm}{Density of universal automata for the samples of \figurename~\ref{fig:median-time}.}}
    \label{fig:density}
 \end{minipage}
\end{figure}

\begin{figure}[!tbp]
    \scalebox{0.5}{\input{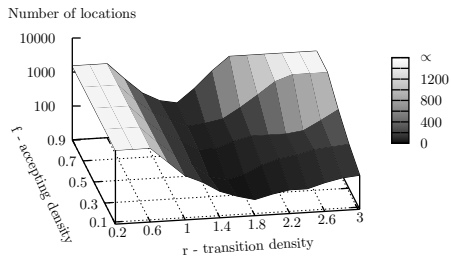}}
    \caption{\protect\parbox[t]{80mm}{Automata size for which the median execution time to
check universality is less than 20 seconds (log scale). See also Table~\ref{tab:max-size}.}}
    \label{fig:max-size} 
\end{figure}

\begin{table}[!tbp]
\caption{Automata size (NBW) for which the median execution time for 
checking universality is less than 20 seconds. The symbol $\propto$ means \emph{more than 1500}.} \label{tab:max-size}
\begin{center}
\begin{tabular}{|c||*{11}{p{5mm}|}p{7mm}|*{3}{p{5mm}|} }\hline
{\scriptsize \backslashbox[0mm]{{\normalsize f}}{{\normalsize r{\strut}}}}
    & \centering 0.2       & \centering 0.4       & \centering 0.6       & \centering 0.8 & \centering 1.0 & \centering 1.2 & \centering 1.4 & \centering 1.6 & \centering 1.8 & \centering 2.0  & \centering 2.2  & \centering 2.4  & \centering 2.6  & \centering 2.8  & \multicolumn{1}{c|}{3.0}  \\\hline\hline
0.1 & \centering $\propto$ & \centering $\propto$ & \centering $\propto$ & \centering 550 & \centering 200 & \centering 120 & \centering 60  & \centering 40  & \centering 30  & \centering 40   & \centering 50   & \centering 50   & \centering 70   & \centering 90   & \multicolumn{1}{c|}{100}  \\\hline
0.3 & \centering $\propto$ & \centering $\propto$ & \centering $\propto$ & \centering 500 & \centering 200 & \centering 100 & \centering 40  & \centering 30  & \centering 40  & \centering 70   & \centering 100  & \centering 120  & \centering 160  & \centering 180  & \multicolumn{1}{c|}{200}  \\\hline
0.5 & \centering $\propto$ & \centering $\propto$ & \centering $\propto$ & \centering 500 & \centering 200 & \centering 120 & \centering 60  & \centering 60  & \centering 90  & \centering 120  & \centering 120  & \centering 120  & \centering 140  & \centering 260  & \multicolumn{1}{c|}{500}  \\\hline
0.7 & \centering $\propto$ & \centering $\propto$ & \centering $\propto$ & \centering 500 & \centering 200 & \centering 120 & \centering 70  & \centering 80  & \centering 100 & \centering 200  & \centering 440  & \centering 1000 & \centering $\propto$ & \centering $\propto$ & \multicolumn{1}{c|}{$\propto$} \\\hline
0.9 & \centering $\propto$ & \centering $\propto$ & \centering $\propto$ & \centering 500 & \centering 180 & \centering 100 & \centering 80  & \centering 200 & \centering 600 & \centering $\propto$ & \centering $\propto$ & \centering $\propto$ & \centering $\propto$ & \centering $\propto$ & \multicolumn{1}{c|}{$\propto$} \\\hline
\end{tabular}
\end{center}
\end{table}

The results are shown in Fig.~\ref{fig:max-size}, and the corresponding 
values are given in Table~\ref{tab:max-size}. The vertical scale is logarithmic.
For example, for $r=2$ and $f=0.5$, our algorithm was able to handle at least $50$ automata of size $120$ 
in less than 20 seconds and was not able to do so for automata of size $140$.
In comparison, Tabakov and Vardi have studied the behavior of
Kupferman-Vardi and Miyano-Hayashi constructions for different
implementation schemes. We compare with the performances of their
symbolic approach which is the most efficient. 
For the same parameter values ($r=2$ and $f=0.5$), they
report that their implementation can handle NBW with at most 8 states 
in less than $20$ seconds~\cite{TabakovV07}.     %%%%%% see page 21 of 29
% For the easier instances
% $r=2.5$ and $f=0.9$, they can analyze automata of size at most $200$ while
% we go over $1500$ states. %%%%%% see page 14 of 29

In \figurename~\ref{fig:scalability}, we show the median execution time to 
check universality for relatively difficult instances ($r=2$ and $f$ vary from
$0.3$ to $0.7$). The vertical scale is logarithmic, so the behavior is roughly 
exponential in the size of the automata. Similar analyzes are reported 
in~\cite{TabakovV07} but for sizes below $10$.

Finally, we give in \figurename~\ref{fig:distribution} the
distribution of execution times for $100$ automata of size $50$ with
$r=2.2$ and $f=0.5$, so that roughly half of the instances are
universal. Each point represents one automaton, and one point lies
outside the figure with an execution time of $675$s for a non
universal automaton. The existence of very few instances that are very
hard was often encountered in the experiments, and this is why we use
the median for the execution times. If we except this hard instance,
\figurename~\ref{fig:distribution} shows that universal automata
(average time $350$ms) are slightly easier to analyze than
non-universal automata (average time $490$ms).  This probably comes
from the fact that we stop the computation of the (greatest) fixed
point whenever the initial state is not in the $\preceq_{{\sf univ}}$-closure
of the current approximation. Indeed, in such
case, since the approximations are $\preceq_{{\sf univ}}$-decreasing,
we know that the initial state would also not lie in the fixed point.
Of course, this optimization applies only for universal automata.

\begin{figure}[!tbp]
  \begin{minipage}[b]{.47\linewidth}
    \scalebox{0.5}{\input{figures/scalability.pstex}}
    \caption{\protect\parbox[t]{40mm}{Median time to check universality (of $100$ automata for each sample point).}}
    \label{fig:scalability}
 \end{minipage} \hfill 
 \begin{minipage}[b]{.47\linewidth}
    \scalebox{0.5}{\input{figures/distribution.pstex}}
    \caption{\protect\parbox[t]{40mm}{Execution time to check universality of 100 automata, 57 of which were universal.}}
    \label{fig:distribution}
 \end{minipage}
\end{figure}

% Other :
% \begin{itemize}
%   \item report on the most difficult instances
%   \item report on the parameter landscape : portion of universal automata 
%     for the parameters {\bf still running for automata of size $20$}.
%   \item report on the meantime, report on the fact that some instances 
%     are very difficult ?
%   \item report more on scalability : other parameters ???
% \end{itemize}

% Section 7

\section{Language Inclusion for B\"uchi automata}\label{sec:inclusion-NBW}
% inclusion

Let $\A_1 = \tuple{\Loc_1, \iota_1, \Sigma, \delta_1, \alpha_1}$ and $\A_2$ 
be two NBW defined on the same alphabet $\Sigma$ for which we want
to check language inclusion: $\L(\A_1) \subseteq^? \L(\A_2)$. To solve
this problem, we check emptiness of $\L(\A_1) \cap \L^c(\A_2)$. 
As we have seen, we can use the Kupferman-Vardi and Miyano-Hayashi construction
to specify a NBW $\A_2^c = \tuple{\Loc_2, \iota_2, \Sigma, \delta_2, \alpha_2}$ 
that accepts the complement of the language of $\A_2$.

Using the classical product construction, let $\B = \A_1 \times \A_2^c$ be a finite automaton
with set of locations $\Loc_{\B} = \Loc_1 \times \Loc_2$, initial state 
$\iota_{\B} = (\iota_1, \iota_2)$, and transition function $\delta_{\B}$ such that 
$\delta_{\B}((\l_1, \l_2), \sigma) = \delta_1(\l_1,\sigma) \times \delta_2(\l_2,\sigma)$.
We equip $\B$ with the generalized B\"uchi condition 
$\{\beta_1,\beta_2\} = \{\alpha_1 \times \Loc_2, \Loc_1 \times \alpha_2\}$, thus asking 
for a run of $\B$ to be accepting that it visits $\beta_1$ and $\beta_2$
infinitely often. It is routine to show that we have $\L(\B) = \L(\A_1) \cap \L(\A_2^c)$. 
The following fixed point
$$\F'_{\B} \equiv \nu y \cdot \Big( 
		\mu x_1 \cdot \big[ \Pre^{\B}(x_1) \cup ( \Pre^{\B}(y) \cap \beta_1 ) \big] \cap
		\mu x_2 \cdot \big[ \Pre^{\B}(x_2) \cup ( \Pre^{\B}(y) \cap \beta_2 ) \big]
	\Big)
$$
\noindent
can be used to check emptiness of $\B$ as we have $\L(\B) \neq \emptyset$ 
iff $\iota_{\B} \in \F'_{\B}$.  
We now define the pre-order $\preceq_{{\sf inc}}$ over the locations of $\B$: 
for all $(\l_1,\l_2),(\l_1',\l_2') \in \Loc_{\B}$, 
let $(\l_1,\l_2) \preceq_{{\sf inc}} (\l_1',\l_2')$ iff $\l_1 = \l'_1$ and 
$\l_2 \preceq_{{\sf univ}} \l_2'$. 
% The next theorem states that $\preceq_{{\sf inc}}$ is a simulation for $\B$.

We extend the definition of simulation relation $\preceq$ (Definition~\ref{def:simulation})
to generalized B\"uchi automata $\B$ by asking that for each $\beta_i$, 
the relation~$\preceq$ is a simulation for $\B$ with accepting states $\beta_i$.

\begin{lem}
The relation $\preceq_{{\sf inc}}$ is a simulation for $\B$.
\end{lem} 

\proof 
First, observe that equality is a simulation relation for $\A_1$. Then, the
first condition of Definition~\ref{def:simulation} is a direct consequence
of the fact that equality (resp. $\preceq_{{\sf univ}}$) is a simulation 
relation for $\A_1$ (resp. for $\A_2^c$), and that $\B = \A_1 \times \A_2^c$ 
is the product of these automata.
Second, it is easy to see that the sets $\beta_1$ and $\beta_2$ are 
$\preceq_{{\sf inc}}$-closed.
\qed

As a consequence of the last lemma, we know that all sets that we
have to manipulate to solve the language inclusion problem using the
fixed point $\F'_{\B}$ are $\preceq_{{\sf inc}}$-closed. The operators
$\cup$, $\cap$ and $\Pre$ can be thus computed efficiently, using the same 
algorithms and data structures as for universality. In particular, let 
$\Pre^{{\sf inc}}_\sigma(\l_1',\l_2') = \Pre^{\A_1}_\sigma(\l_1')
 \times
 \Pre^{{\sf univ}}_\sigma(\l_2')
$
where
$\Pre^{{\sf univ}}_{\sigma}$ is computed by Algorithm~2 (with input $\A_2$).
It is easy to show as a corollary of Theorem~\ref{theo:correctness-alg-pre} that 
$\closure{\Pre^{{\sf inc}}_\sigma(\l_1',\l_2')} = \Pre^{\B}_{\sigma}(\closure{\{(\l_1',\l_2')\}})$.

% Section 8
\section{Conclusion}\label{sec:conclusion}

We have shown that the prohibitive complementation constructions
for nondeterministic B\"uchi automata can be avoided for solving 
classical problems like universality and language inclusion. 
Our approach is based on fixed points computation and the existence of simulation relations 
for the (exponential) constructions used in complementation of 
B\"uchi automata. Those simulations are used to dramatically 
reduce the amount of computations needed to decide classical problems.
Their definition relies on the structure of the original automaton
and do not require explicit complementation. 

The resulting algorithms evaluate a fixed point formula and avoid
redundant computations by maintaining sets of maximal elements
according to the simulation relation. In practice, the computation of
the predecessor operator, which is the key of the approach, is
efficient because it is done on antichains of elements only. Even though
the classical approaches (as well as ours) have the same worst case
complexity, our prototype implementation outperforms those approaches
where the structural properties of the complement automaton (witnessed by the 
existence of simulation relations) is not exploited. The huge gap of performances holds
over the entire parameter space of the randomized model proposed by
Tabakov and Vardi.

Applications of this paper go beyond universality and language inclusion 
for NBW, as we have shown that the methodology applies to alternating 
B\"uchi automata for which efficient translations from LTL formula are 
known~\cite{GastinOddoux}. Significant improvements in the model-checking 
and satisfiability problem of LTL can be achieved with the same ideas~\cite{DDMR08a,DDMR08c}.

\paragraph{{\bf Acknowledgment.}} We thank two anonymous reviewers for helpful 
comments and suggestions.

\bibliography{biblio}
\bibliographystyle{alpha}

\iffalse

\fi

% \input{appendix.tex}

\end{document}